\documentclass[reprint]{JASAnew}
\usepackage{xcolor}

\begin{document}

\title[]{Wave propagation in an elastic lattice with non-reciprocal stiffness and engineered damping}

\author{Harshit Kumar Sandhu}
\author{Saurav Dutta}
\author{Rajesh Chaunsali}
\email{Email: rchaunsali@iisc.ac.in}
\affiliation{Department of Aerospace Engineering,  Indian Institute of Science, Bangalore 560012, India}

\preprint{Author, JASA}		

\date{\today}

\begin{abstract}
Nonreciprocal wave propagation allows for directional energy transport. In this work, we systematically investigate wave dynamics in an elastic lattice that combines nonreciprocal stiffness with viscous damping. After establishing how conventional damping counteracts the system's gain, we introduce a non-dissipative form of nonreciprocal damping in the form of gyroscopic damping. We find that the coexistence of nonreciprocal stiffness and nonreciprocal damping results in a decoupled control mechanism. The nonreciprocal stiffness is shown to govern the temporal amplification rate, while the nonreciprocal damper independently tunes the wave's group velocity and oscillation frequency. This decoupling gives rise to phenomena such as the enhancement of net amplification for slower-propagating waves, and also boundary-induced wave {interference arising from divergent and convergent reflected wave trajectories with varying growth rates.} These findings provide a theoretical framework for designing active metamaterials with more versatile control over their wave propagation characteristics.

\end{abstract}


\maketitle

\section{\label{sec:1}Introduction}
The ability to control the flow of wave energy is a central theme in physics and engineering. Systems that break reciprocity—the symmetry of transmission between a source and a receiver—enable unique wave phenomena and have led to applications such as acoustic diodes, unidirectional amplifiers, and robust vibration isolators~\citep{Nassar2020}. A common method to achieve nonreciprocity is to engineer asymmetric couplings in a lattice, an idea rooted in the Hatano-Nelson model~\citep{H-N-1996, Hatano-1998}. This approach creates active systems where waves are amplified in one direction while being attenuated in the opposite~\citep{Brandenbourger2019}. This concept has more recently been placed within the broader context of non-Hermitian topology~\citep{Liang2018, Yao-2018}, which connects nonreciprocal couplings to the non-Hermitian skin effect—the localization of bulk modes at a system's boundaries~\citep{Ding2022, Ching2023}.

The physical realization of these models often relies on active feedback control, where sensors and actuators generate direction-dependent interaction forces. Such nonreciprocal couplings have been demonstrated on various platforms, including acoustic~\citep{Baile2021, Maddi2024}, mechanical~\citep{Brandenbourger2019, Ghatak2020, Guancong2022}, and electric lattices~\citep{Zhang2021, jana2025harnessingnonlinearitytamewave}. Since momentum is locally injected by each element to amplify bulk waves in a preferential direction, these lattices are distinct from momentum-conserving ``odd elastic" solids~\citep{Scheibner2020, Zhao2020, Banerjee2021, Chen2021, Gao2022}.

Despite these advances, research has predominantly focused on nonreciprocal stiffness, where interaction forces are proportional to relative displacements~\citep{Brandenbourger2019, rosa2020dynamics}. In this context, damping is generally treated as a parasitic effect that simply counteracts the system's gain. This perspective, however, overlooks the possibility of engineering damping itself as a functional tool for wave manipulation.

In this paper, we explore this possibility by introducing a nonreciprocal damping mechanism into a lattice with nonreciprocal stiffness. We demonstrate that this interplay gives rise to a decoupled control mechanism. Specifically, we show that the nonreciprocal stiffness governs the wave's temporal amplification rate, while the nonreciprocal damping (gyroscopic damping \cite{Carta2014}) term independently tunes its group velocity and oscillation frequency. This decoupling offers a more versatile method for controlling wave propagation in active media than is achievable with nonreciprocal stiffness alone. 

The paper is organized as follows. Section~\ref{sec:2} reviews the model with nonreciprocal stiffness and analyzes its interaction with conventional damping. In Sec.~\ref{sec:3}, we introduce a specialized, non-dissipative, nonreciprocal damper and examine its isolated effect on wave propagation. The full system combining both types of nonreciprocity is investigated in Sec.~\ref{sec:4}, {where we detail the decoupled control mechanism and its consequence on tailoring the net amplification. We further demonstrate boundary-induced wave interference, showing how the interplay between stiffness and damping asymmetries can be engineered to combine frequency components.} Finally, Sec.~\ref{sec:5} summarizes our findings and concludes the paper.

\begin{figure}[!]
\includegraphics[width=\reprintcolumnwidth]{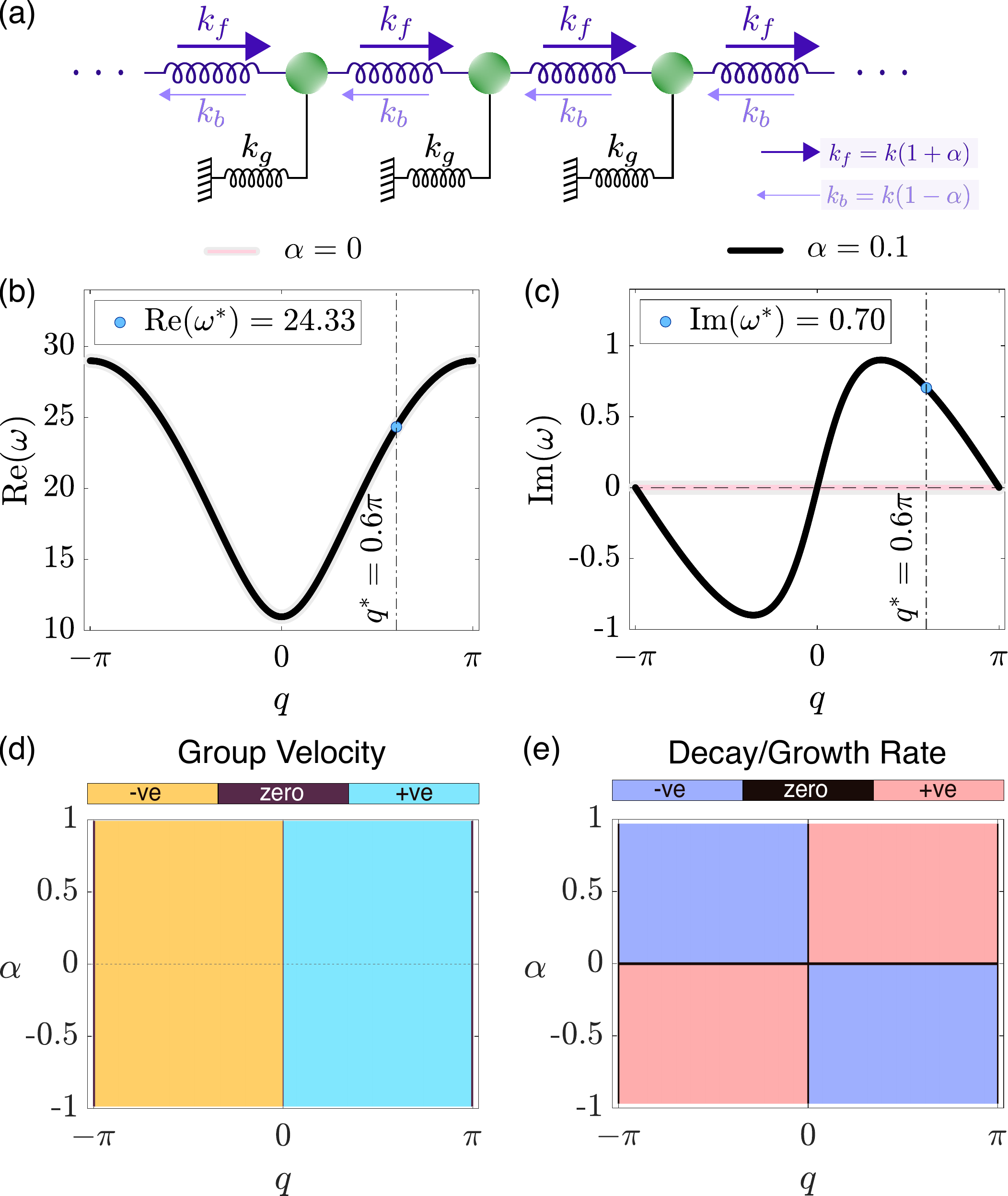}
\caption{\label{fig:FIG1}{\textbf{Wave propagation in a 1D lattice with nonreciprocal stiffness.} 
(a) Schematic of the lattice with asymmetric intersite springs characterized by forward and backward stiffnesses \(k_f = k(1+\alpha)\) and \(k_b = k(1-\alpha)\). 
(b–c) Real and imaginary parts of the complex dispersion relation \(\omega(q)\) for asymmetry parameter \(\alpha = 0\) and \(\alpha = 0.1\).  
(d–e) Parametric maps of group velocity and decay/growth rate signs as functions of wavenumber \(q\) and asymmetry \(\alpha\), highlighting directional energy transport in the absence of damping.
We take intersite stiffness \(k = 180\), onsite stiffness \(k_g = 120\), and mass \(m = 1\) in these calculations.}}
\raggedright
\end{figure}

\section{\label{sec:2}Lattice with nonreciprocal stiffness}

We consider a one-dimensional elastic lattice as illustrated in Fig.~\ref{fig:FIG1}(a). Each unit cell contains a point mass \(m\), attached to an onsite (ground) spring of stiffness \(k_g\), and coupled to neighboring masses via nonreciprocal intersite springs. The stiffness in the forward direction is \(k_f = k(1+\alpha)\), and in the backward direction is \(k_b = k(1-\alpha)\), where \(\alpha \in [-1,1]\) quantifies the degree of nonreciprocity.

In the following subsections, we analyze wave propagation characteristics in such lattices, beginning with the undamped case and subsequently incorporating the effects of viscous damping.

\subsection{\label{subsec:2:1} No damper}

We first revisit the lattice without any damper, which has been addressed in several prior studies~\cite{Brandenbourger2019}. The equation of motion for the displacement \(u_n(t)\) of the \(n\)th mass is given by
\begin{equation}
m \ddot{u}_{n} + (2k + k_{g}) u_{n} - k(1+\alpha) u_{n-1} - k(1-\alpha) u_{n+1} = 0.
\end{equation}

To analyze wave propagation, we employ the Bloch-Floquet ansatz \(u_{n}(t) = \hat{u} e^{i(qn - \omega t)}\), where \(\hat{u}\) is the amplitude, \(q\) is the wavenumber, and \(\omega\) is the angular frequency. Substituting into the equation of motion yields the dispersion relation:
\begin{equation}
\omega(q) = \sqrt{\frac{1}{m} \left[ k_g + 2k (1 - \cos q) + 2ik\alpha\sin q \right]}.
\end{equation}

For \(\alpha \neq 0\), the dispersion relation yields complex-valued \(\omega\) for real \(q\). This implies both temporal oscillations, captured by \(\text{Re}[\omega(q)]\), and temporal growth or decay, governed by \(\text{Im}[\omega(q)]\). The real and imaginary components can be expressed as
\begin{align}
\text{Re}[\omega(q)] &= \sqrt{r(q)} \cos\left(\frac{\theta(q)}{2}\right), \label{eq:EQ5} \\
\text{Im}[\omega(q)] &= \sqrt{r(q)} \sin\left(\frac{\theta(q)}{2}\right), \label{eq:EQ6}
\end{align}
where the auxiliary quantities are defined as
\[
r(q) = \sqrt{A(q)^2 + B(q)^2}, \quad \theta(q) = \operatorname{atan2}\left(B(q), A(q)\right),
\]
\[
A(q) = \frac{1}{m} \left[ k_g + 2k (1 - \cos q) \right], \quad B(q) = \frac{2k\alpha \sin q}{m}.
\]

Figures~\ref{fig:FIG1}(b) and \ref{fig:FIG1}(c) show the real and imaginary parts of the dispersion relation for \(\alpha = 0.1\), alongside the reciprocal case (\(\alpha = 0\)) for comparison {(we choose $k = 180$ and $k_g = 120$ as representative values for all simulations, as they clearly illustrate the phenomena and also yield operating frequencies on the same order of magnitude as recent experiments~\cite{Brandenbourger2019})}. 
The real part, \(\text{Re}[\omega(q)]\), exhibits negligible variation, whereas the imaginary part, \(\text{Im}[\omega(q)]\), reveals significant nonreciprocal behavior: attenuation for \(q < 0\) and amplification for \(q > 0\). These behaviors become clearer in the small-\( \alpha \) limit, where the dispersion relation admits the following approximations:
\begin{align}
\text{Re}[\omega(q)] &= \sqrt{A(q)} + \frac{k^2 \alpha^2 \sin^2 q}{2 m^2 A(q)^{3/2}} + \mathcal{O}(\alpha^3), \\
\text{Im}[\omega(q)] &= \frac{k\alpha \sin q}{m \sqrt{A(q)}} + \mathcal{O}(\alpha^3).
\end{align}
These expressions reveal that the leading-order correction appears exclusively in the imaginary part of the complex frequency, $\text{Im}[\omega(q)]$. Consequently, this term has a dominant effect on the temporal amplification or decay of the wave, while the corresponding correction to the real part of the frequency is of a higher order. 
{It is noteworthy that this formulation results in an \textit{increase} in the oscillation frequency for both positive and negative values of $\alpha$. This behavior contrasts with the findings of Rosa and Ruzzene~\cite{rosa2020dynamics}, where a different stiffness definition ($k_f = k(1-\alpha)$ and $k_b = k$) was employed. Their definition leads to a \textit{decrease} in frequency for positive $\alpha$ and an \textit{increase} for negative $\alpha$.}

We also observe that \(\text{Re}[\omega(q)]\) is symmetric about \(q = 0\), while \(\text{Im}[\omega(q)]\) is antisymmetric. This property holds for all values of \(\alpha\), as \(r(q)\) is an even function and \(\theta(q)\) is an odd function of \(q\), thereby imparting the observed symmetry properties to the dispersion relations in Eqs.~\eqref{eq:EQ5} and \eqref{eq:EQ6}.

\begin{figure}[!]
\includegraphics[width=\reprintcolumnwidth]{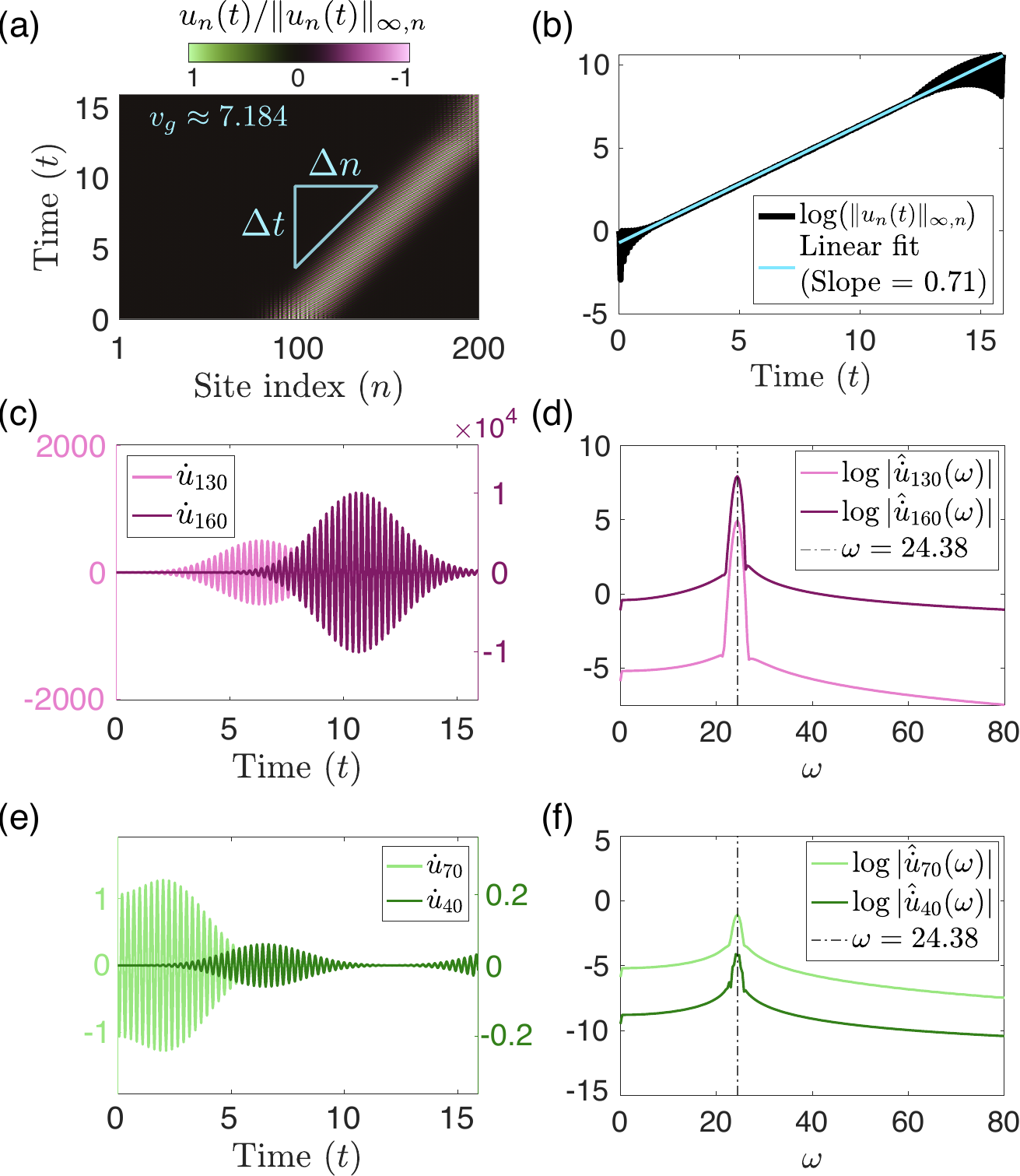}
\caption{\label{fig:FIG2}{\textbf{{Finite-chain simulations of wave propagation and directional amplification in a lattice with nonreciprocal stiffness.}}
{(a) Space–time evolution of the space-normalized displacement field, $u_n(t)/\|u_n(t)\|_{\infty,n}$, in a 200-particle lattice with fixed boundaries showing selective amplification in the forward direction.
The wavefront travels with group velocity \( v_g \approx 7.184 \).}
(b) Logarithmic plot of the instantaneous global maximum of the lattice displacement field, with a linear fit yielding a slope of 0.71. 
(c–d) Velocity responses and corresponding spectral amplitudes at downstream sites (\(n = 130, 160\)) highlight energy amplification and a {spectral peak} near \( \omega \approx 24.38 \, \text{rad/s} \). 
(e–f) Upstream sites {(\(n = 70, 40\))} exhibit strong attenuation in both time and frequency domains, confirming unidirectional energy transport enabled by non-reciprocal stiffness.}}
\raggedright
\end{figure}

These observations imply that a wave packet centered at a particular frequency will travel symmetrically in both directions with identical group velocities, given by \(\frac{d}{dq} \, \text{Re}[\omega(q)]\), while undergoing asymmetric amplification or decay.  Figures~\ref{fig:FIG1}(d) and \ref{fig:FIG1}(e) present parametric plots showing the sign of the group velocity and the amplification/decay rates as functions of \(\alpha\). The group velocity maintains its sign across all values of \(\alpha\), with zeros at \(q = 0, \pm\pi\). The growth/decay map indicates that amplification occurs when \(\alpha\) and \(q\) share the same sign. Reversing the sign of \(\alpha\) reverses the direction of amplification.

Next, we perform numerical simulations and verify the nonreciprocal wave propagation in a finite lattice setting shown in Fig.~\ref{fig:FIG2}. We take a 200-particle chain with fixed boundary conditions and provide an initial condition in the middle of the chain in the form of a Gaussian-modulated wave packet centered at \( |q^*| = 0.6\pi \), corresponding to a temporal growth [marked in Fig.~\ref{fig:FIG1}(c)]. The prescribed initial displacement (with vanishing velocity) of the $n$th particle is given by $$ u_n(0) =  \exp\left[-\frac{(n - n_0)^2}{2\sigma_n^2}\right] {{\cos(q^* (n-n_0))}}, $$
where \( n_0 = 100 \) is the center of the excitation, and \( {\sigma_n} = 12 \) is the envelope width. 

Figure~\ref{fig:FIG2}(a) shows the space-time evolution of the normalized displacement field, {\(u_n(t)/\|u_n(t)\|_{\infty,n}\)}, 
where {\(\|u_n(t)\|_{\infty,n} = \max_n |u_n(t)|\) denotes the spatial maximum of the displacement magnitude at time $t$}. An initial wave packet splits into two components traveling in opposite directions with the same group velocity. However, only the rightward-propagating component is amplified, a clear signature of the system's non-reciprocal behavior. The group velocity measured from the slope of the {forward} propagating wavefront is $v_g \approx 7.184$, which is in close agreement with the slope calculated from the dispersion relation shown in Fig.~\ref{fig:FIG1}(b).

To quantify the one-way amplification, in Fig.~\ref{fig:FIG2}(b), we show the temporal evolution of {log(\(\|u_n(t)\|_{\infty,n}\))}. The envelope of this logarithmic signal exhibits a linear trend, indicating exponential growth in the underlying displacement. A linear fit to the data yields a slope of approximately 0.71, representing the growth rate. This value closely matches the imaginary part of the complex dispersion relation, $\text{Im}[\omega(q^*)] \approx 0.70$, shown in Fig.~\ref{fig:FIG1}(c).

Figure~\ref{fig:FIG2}(c) shows the velocity time series for particles at positions $n=130$ and $n=160$, which are downstream from the excitation point. Both signals exhibit clear temporal amplification. The corresponding frequency spectra, presented in Fig.~\ref{fig:FIG2}(d), reveal a {spectral} peak at $\omega \approx 24.38$. This value is consistent with the oscillation frequency, $\text{Re}[\omega(q^*)]$, from Fig.~\ref{fig:FIG1}(b), {which corresponds to the central frequency of our narrow bandwidth Gaussian packet.}

In contrast, the velocity signals from upstream positions ($n=70$ and $n=40$), plotted in Fig.~\ref{fig:FIG2}(e), show significant decay over time. The associated frequency spectra in Fig.~\ref{fig:FIG2}(f) contain the same {spectral peak}, but its amplitude is substantially reduced. This confirms that while the initial disturbance propagates in both directions, non-reciprocity leads to strong wave amplification in the forward (downstream) direction and attenuation in the backward (upstream) direction.

\begin{figure}[!]
\includegraphics[width=\reprintcolumnwidth]{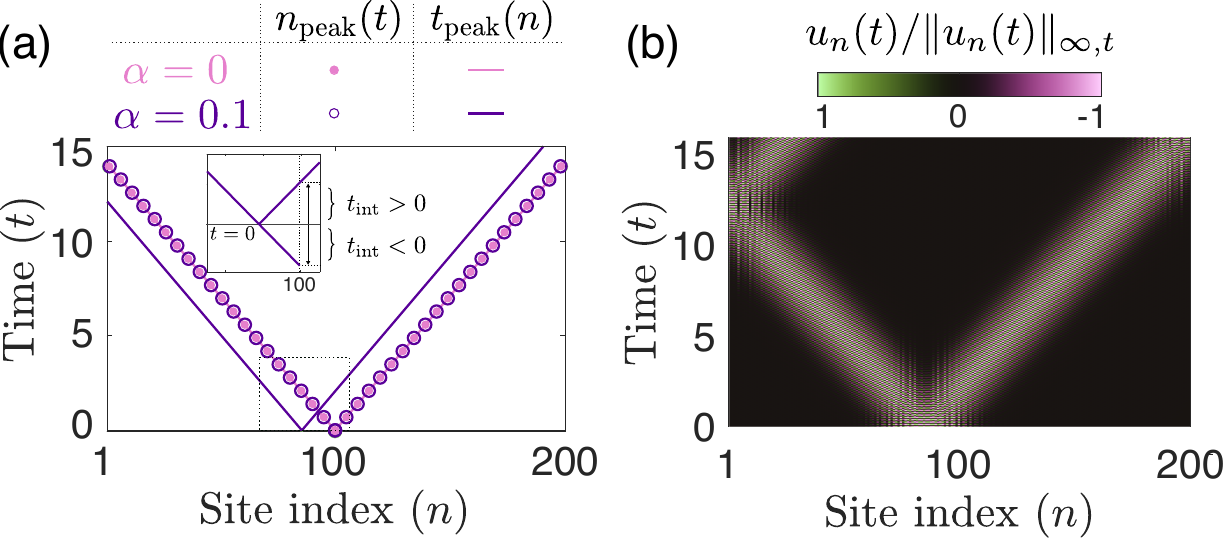}
\caption{\label{fig:FIG3}
{{\textbf{Distinct trajectories of spatial and temporal peaks in a 1D finite lattice with nonreciprocal stiffness.} (a) Analytical trajectories of the Gaussian envelope maxima, comparing the instantaneous spatial peaks $n_{\mathrm{peak}}(t)$ (dots and circles) and the temporal peaks $t_{\mathrm{peak}}(n)$ (solid lines). While the forward and backward $t_{\mathrm{peak}}(n)$ branches share the same slope (indicating equal effective group velocity), they exhibit distinct temporal intercepts at the source $n=n_0$, as highlighted in the inset. (b) Spatiotemporal evolution of the displacement field normalized by the local temporal maximum, $u_n(t)/\|u_n(t)\|_{\infty,t}$. The wavefronts propagate with equal speeds but project to unequal origin times, visually confirming the positive and negative intercepts predicted in (a).}}}
\raggedright
\end{figure}

{{It is interesting to note in Figs.~\ref{fig:FIG2}(c) and \ref{fig:FIG2}(e) that the peak of the Gaussian time history occurs at different time instances, even though the observation points are located spatially symmetrically (e.g., equidistant upstream and downstream) from the initial excitation.
This phenomenon is a specific feature of non-Hermitian systems governed by a complex dispersion relation.

For a narrow-bandwidth Gaussian wave packet with spatial variance $\sigma_n$, the trajectory of the \textit{spatial peak}, $n_{\text{peak}}(t)$, is determined solely by the real part of the group velocity (see Appendix A for derivation):
\begin{equation}
n_{\text{peak}}(t) = n_0 + v_g\,t,
\label{eq:Main3_new}
\end{equation}
where $v_g = \mathrm{Re}[\omega'(q^*)]$ denotes the standard group velocity evaluated at the carrier wavenumber $q^*$.

However, the evolution of the \textit{temporal peak}, $t_{\text{peak}}(n)$—which corresponds to the maximum signal recorded by a stationary observer—depends on both the real and imaginary components of the dispersion. As derived in Appendix A, this relationship is given by:
\begin{equation}
t_{\text{peak}}(n) = \frac{(n-n_0)}{v_g^{\text{eff}}} + t_{\text{int}} \neq \frac{(n-n_0)}{v_g},
\label{eq:Main4_new}
\end{equation}
where the effective velocity is $v_g^{\text{eff}} = \mathrm{Re}[\omega'] - (\mathrm{Im}[\omega'])^2/\mathrm{Re}[\omega']$, and the intercept term is defined as:
\begin{equation}
t_{\text{int}} = \frac{\sigma_n^2\,\mathrm{Im}[\omega(q^*)]}
{\big(\mathrm{Re}[\omega'(q^*)]\big)^2 - \big(\mathrm{Im}[\omega'(q^*)]\big)^2}.
\label{eq:Main5_new}
\end{equation}
Consequently, in a system with asymmetric complex dispersion, the forward and backward propagating packets may share the same group velocity $v_g^{\text{eff}}$ but possess different growth rates $\mathrm{Im}[\omega(q^*)]$.
For instance, as illustrated in Fig.~\ref{fig:FIG1}(c), the forward-propagating wave (positive $q^*$) exhibits a positive growth rate ($\mathrm{Im}[\omega] > 0$), which dictates a positive temporal intercept $t_{\text{int}} > 0$.
Conversely, the backward-propagating wave (negative $q^*$) is subject to attenuation ($\mathrm{Im}[\omega] < 0$), leading to a negative intercept $t_{\text{int}} < 0$.
This difference, as visualized in Fig.~\ref{fig:FIG3}(a), causes the temporal peaks at spatially symmetric locations to occur at different times.
This behavior is further verified in Fig.~\ref{fig:FIG3}(b), which plots the spatiotemporal map with time histories at each site normalized by their respective temporal peak displacement.

To avoid ambiguity between amplification and arrival times, the remaining sections of this work will focus on spatiotemporal maps showing the \textit{spatially normalized} wave packet evolution.
This approach isolates the transport characteristics, allowing us to relate the wave dynamics directly to the real group velocity $v_g$.}}

Thus far, we have focused on systems with only elastic nonreciprocity. We now examine how the introduction of viscous damping influences this behavior. The simplest approach is to incorporate a damper in two ways: (i) onsite damping, which acts on the velocity of individual particles, and (ii) intersite damping, which depends on the relative velocity between neighboring particles. We discuss these cases sequentially in the following sections.

\subsection{\label{subsec:2:2} Onsite Damper}
\begin{figure}[!]
\includegraphics[width=\reprintcolumnwidth]{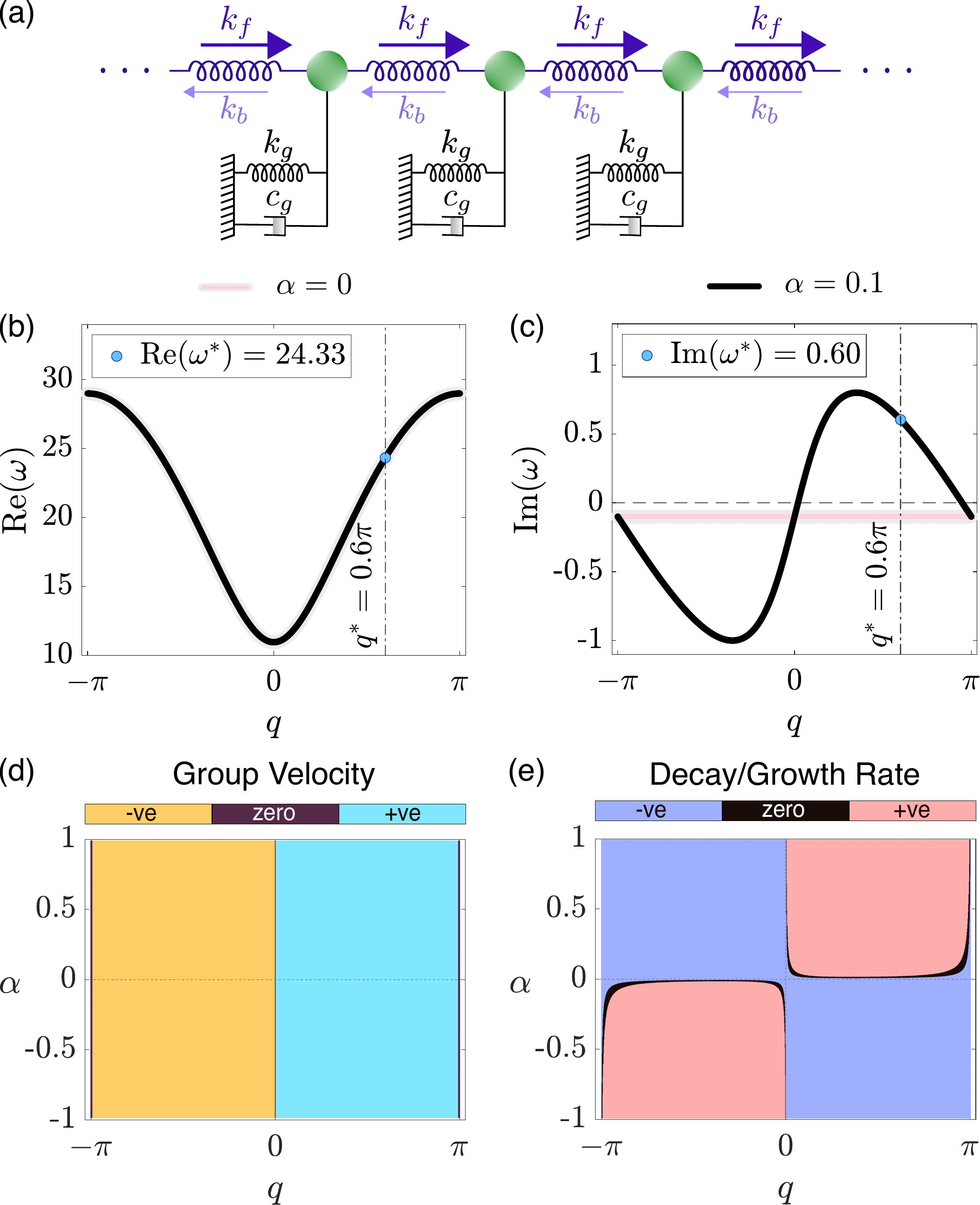}
\caption{\label{fig:FIG4}{\textbf{Wave propagation in a 1D lattice with nonreciprocal stiffness and onsite damping.} 
(a) Schematic of the lattice with asymmetric intersite springs and onsite viscous dampers; 
(b–c) Real and imaginary parts of the complex dispersion relation \(\omega(q)\) for asymmetry parameters \(\alpha = 0\) and \(\alpha = 0.1\). 
(d–e) Parametric maps showing the signs of group velocity and temporal growth/decay rate as functions of \(q\) and \(\alpha\). We take onsite damping \(c_g = 0.2\). Other parameters are the same as in Fig.~\ref{fig:FIG1}.}}
\raggedright
\end{figure}
We now incorporate onsite damping into the lattice model, characterized by the damping coefficient \(c_g\), as shown in Fig.~\ref{fig:FIG4}(a). The equation of motion becomes
\begin{equation}
m \ddot{u}_{n} + c_g \dot{u}_{n} + (2k + k_{g}) u_{n}  
- k(1+\alpha) u_{n-1} - k(1-\alpha) u_{n+1} = 0. \label{eq:EQ8}
\end{equation}
Applying the Bloch-Floquet ansatz yields the following dispersion relation:
\begin{equation}
\omega^2 + i\frac{c_g}{m}\omega - \frac{1}{m} \Big[ k_g + 2k(1 - \cos q) + 
2ik\alpha \sin q \Big] = 0. \label{eq:EQ9}
\end{equation}
Solving this quadratic equation for \(\omega(q)\), we obtain
\begin{equation}
\omega(q) = -i\frac{c_g}{2m} + \sqrt{r(q)} e^{i \theta(q)/2}.
\end{equation}
We select the root with the positive sign to ensure a positive oscillation frequency, corresponding to forward time evolution. Accordingly, the real and imaginary parts are given by
\begin{align}
\text{Re}[\omega(q)] &= \sqrt{r(q)} \cos\left(\frac{\theta(q)}{2}\right), \label{eq:EQ13} \\
\text{Im}[\omega(q)] &= -\frac{c_g}{2m} + \sqrt{r(q)} \sin\left(\frac{\theta(q)}{2}\right), \label{eq:EQ14}
\end{align}
where
\begin{align*}
r(q) &= \sqrt{A(q)^2 + B(q)^2}, \quad \theta(q) = \operatorname{atan2}\left(B(q), A(q)\right), \\
A(q) &= -\left(\frac{c_g}{2m}\right)^2 + \frac{1}{m} \left[ k_g + 2k(1 - \cos q) \right], \\
B(q) &= \frac{2k\alpha \sin q}{m}.
\end{align*}

The dispersion relation for this system is plotted in Figs.~\ref{fig:FIG4}(b) and \ref{fig:FIG4}(c). For small non-reciprocity ($\alpha \sim \mathcal{O}(\epsilon)$)  and damping ($c_g \sim \mathcal{O}(\epsilon)$), it can be approximated by 
\begin{align}
\text{Re}[\omega(q)] &= \sqrt{A_0(q)} + \frac{k^2 \alpha^2 \sin^2 q}{2m^2 A_0(q)^{3/2}} \nonumber \\
&\quad - \frac{c_g^2}{8m^2 \sqrt{A_0(q)}} + \mathcal{O}(\epsilon^3), \label{eq:damped_approx_Real} \\
\text{Im}[\omega(q)] &= \frac{k\alpha \sin q}{m \sqrt{A_0(q)}} -\frac{c_g}{2m} + \mathcal{O}(\epsilon^3), \label{eq:damped_approx_Imag}
\end{align}
where $A_0(q) = \frac{1}{m} \left[ k_g + 2k(1 - \cos q) \right ]$. In contrast to the case with no damper, the imaginary part of the frequency is uniformly {reduced by $c_g/(2m)$ without any dependency on $q$}, as seen in Fig.~\ref{fig:FIG4}(c). Furthermore, the higher-order correction to the real part is now negative, causing a \textit{decrease} in the oscillation frequency, which is opposite to the behavior without a damper. While $\text{Re}[\omega(q)]$ remains a symmetric function of $q$, the constant negative shift renders $\text{Im}[\omega(q)]$ asymmetric.

The parametric plots in Figs.~\ref{fig:FIG4}(d) and \ref{fig:FIG4}(e) illustrate the interplay between non-reciprocity and damping. While the sign of group velocity is largely unchanged, the boundaries in the growth/decay map are significantly altered, leading to a reduction in the parameter space for amplification. A notable feature is the existence of traveling waves ($\text{Im}[\omega(q)] = 0$) at wavenumbers away from $q=0$ and $q=\pm\pi$. By setting Eq.~\eqref{eq:damped_approx_Imag} to zero in the long-wavelength limit ($q\to 0$), we can find the condition for these waves:
\begin{equation}
\alpha q \approx \frac{c_g}{2k} \sqrt{\frac{k_g}{m}}.
\label{eq:hyperbolic_condition}
\end{equation}
This hyperbolic relationship between $\alpha$ and $q$ for neutrally stable traveling waves can be verified by the boundary line separating the gain and decay regions near the origin in Fig.~\ref{fig:FIG4}(e).

\begin{figure}[!]
\includegraphics[width=\reprintcolumnwidth]{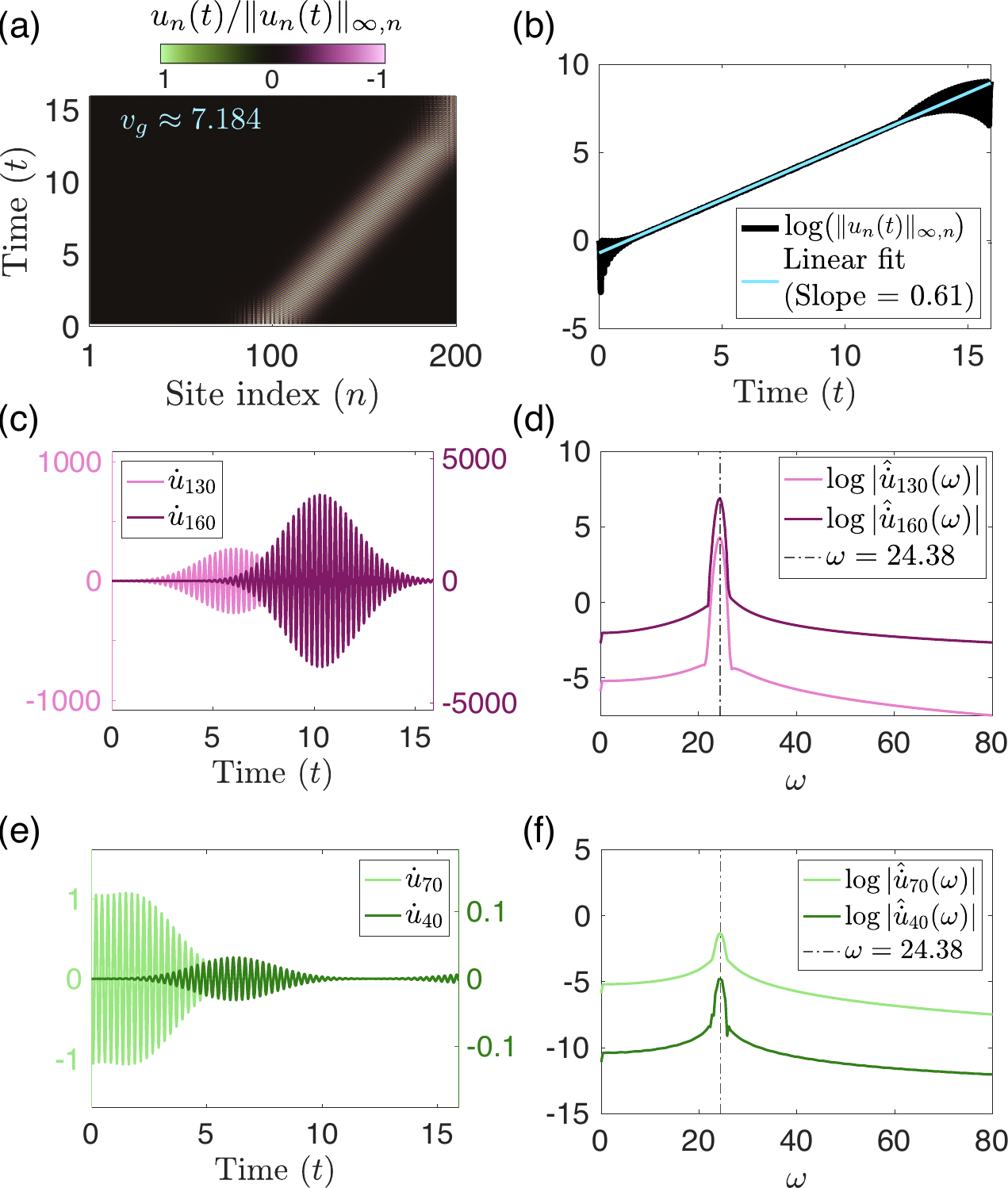}
\caption{\label{fig:FIG5}{{\textbf{Finite-chain simulations of wave propagation and directional amplification in a lattice with nonreciprocal stiffness and onsite damping.} 
(a) Space–time evolution of the space-normalized displacement field, showing amplification in the forward-propagating wave packet} with measured group velocity \( v_g \approx 7.184 \).
(b) Logarithmic plot of the instantaneous global maximum, where a linear fit yields a slope of 0.61, indicating a reduced temporal growth rate due to the presence of onsite damping. 
(c–d) Velocity responses and corresponding spectral amplitudes at downstream sites (\(n = 130, 160\)) confirm energy amplification and {peak} excitation near \( \omega \approx 24.38 \). 
(e–f) Upstream sites {(\(n = 70, 40\))} show significant attenuation in both time and frequency domains, affirming unidirectional energy transport.}}
\raggedright
\end{figure}


We now examine the effect of onsite damping on the wave dynamics of a finite lattice, with the results presented in Fig.~\ref{fig:FIG5}. The initial perturbation is a spatially localized Gaussian displacement near the center of the lattice, identical to the undamped case. While both forward and backward propagating waves are observed, their amplitudes evolve differently due to the interplay between nonreciprocal stiffness and damping.

Figure~\ref{fig:FIG5}(a) shows unidirectional wave propagation with a forward group velocity of ${v_{g}} \approx 7.18$, indicating that a small onsite damping does not significantly alter the wave speed. From Fig.~\ref{fig:FIG5}(b), we measure a growth rate of $0.61$, which confirms that the amplification persists, albeit at a reduced rate. The growth of velocity amplitudes at a downstream location, depicted in Figs.~\ref{fig:FIG5}(c) and \ref{fig:FIG5}(d), further confirms this unidirectional amplification. This effect is weaker than in the undamped case; however, the spectral peak at $\omega \approx 24.38$ remains largely unaffected. Conversely, the backward-propagating signal shown in Figs.~\ref{fig:FIG5}(e) and \ref{fig:FIG5}(f) is more strongly attenuated than its undamped counterpart, while its {peak} frequency is unchanged. 


Overall, these findings demonstrate that the nonreciprocal amplification mechanism and its kinematic signatures survive in the presence of a small onsite damping.

\subsection{\label{subsec:2:3} Intersite Damper}
\begin{figure}[!]
\includegraphics[width=\reprintcolumnwidth]{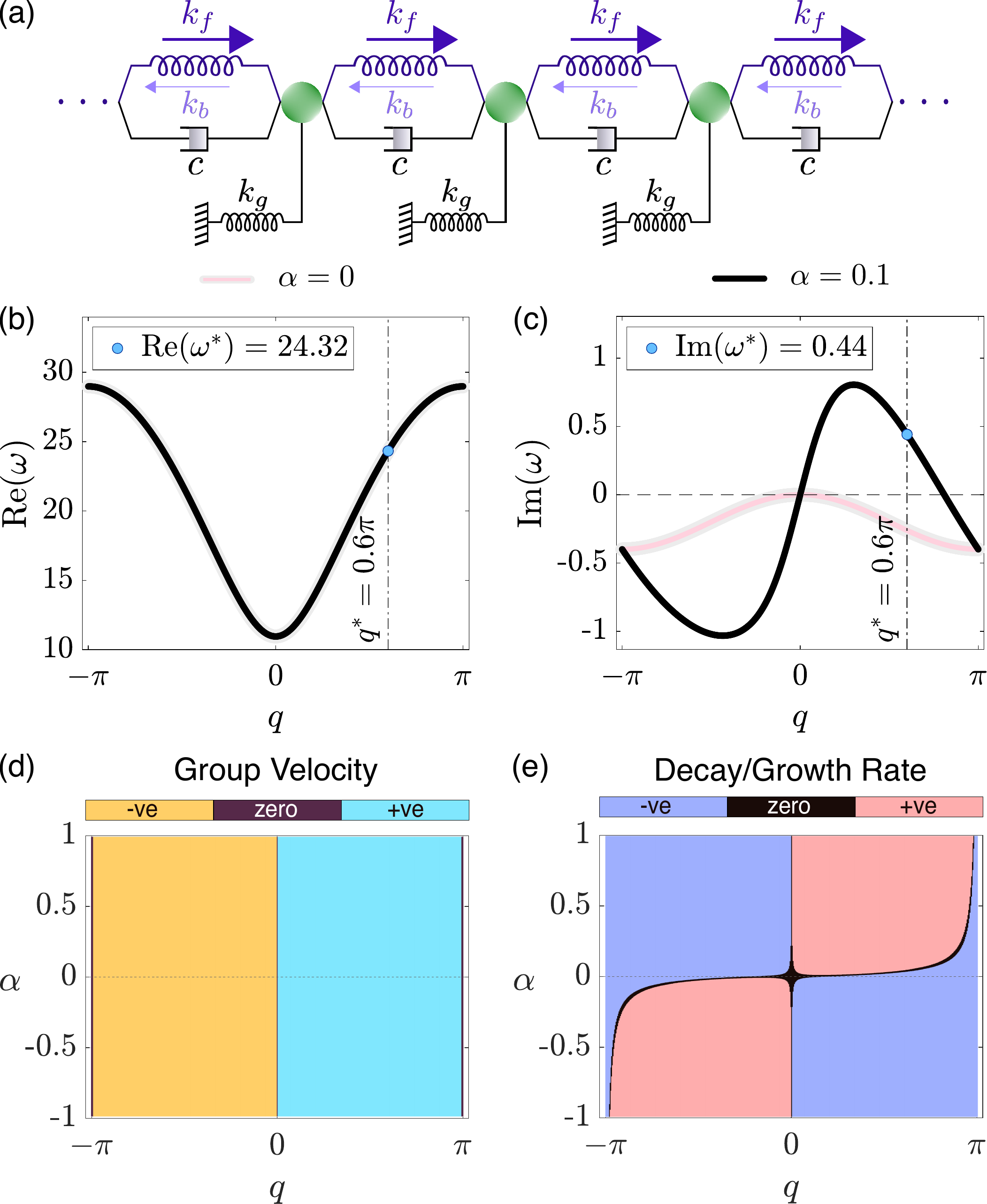}
\caption{\label{fig:FIG7}{\textbf{Wave propagation in a 1D lattice with nonreciprocal stiffness and intersite damping.} 
(a) Schematic of the lattice featuring direction-dependent springs and intersite viscous dampers. 
(b–c) Real and imaginary parts of the complex dispersion relation \( \omega(q) \) for \( \alpha = 0 \) and \( \alpha = 0.1 \).
(d–e) Parametric plots of the signs of group velocity and decay/growth rate as functions of \( q \) and \( \alpha \).
We take \( c = 0.2 \), and keep other parameters the same as in Fig.~\ref{fig:FIG1}. }}
\raggedleft
\end{figure}
We now introduce intersite damping into the lattice model, characterized by the damping coefficient $c$, as shown in Fig.~\ref{fig:FIG7}(a). The equation of motion becomes
\begin{equation}
\begin{split}
    m \ddot{u}_{n} &+ c(2\dot{u}_{n} - \dot{u}_{n-1} - \dot{u}_{n+1}) \\
    &+ (2k + k_{g}) u_{n} 
    - k(1+\alpha) u_{n-1} - k(1-\alpha) u_{n+1} = 0.
\end{split}
\label{eq:EOM_intersite}
\end{equation}
Applying the Bloch-Floquet ansatz yields the following dispersion relation:
\begin{equation}
\begin{split}
\omega^2 &+ i\frac{2c(1 - \cos q)}{m} \omega \\
&- \frac{1}{m} \left[ k_g + 2k(1 - \cos q) + 2ik\alpha\sin q \right] = 0.
\end{split}
\label{eq:char_eq_intersite}
\end{equation}

Solving this quadratic equation for $\omega(q)$, we obtain
\begin{equation}
\omega(q) = -i\frac{c(1-\cos q)}{m} + \sqrt{r(q)} e^{i \theta(q)/2}.
\end{equation}
We select the root with the positive sign to ensure a positive oscillation frequency, corresponding to forward time evolution. Accordingly, the real and imaginary parts are given by
\begin{align}
\text{Re}[\omega(q)] &= \sqrt{r(q)} \cos\left(\frac{\theta(q)}{2}\right), \label{eq:real_intersite}\\
\text{Im}[\omega(q)] &= -\frac{c}{m}(1 - \cos q) + \sqrt{r(q)} \sin\left(\frac{\theta(q)}{2}\right), \label{eq:imag_intersite}
\end{align}
where
\begin{align*}
r(q) &= \sqrt{A(q)^2 + B(q)^2}, \quad \theta(q) = \operatorname{atan2}\left(B(q), A(q)\right), \\
A(q) &= \frac{1}{m}\left[k_g + 2k(1-\cos q)\right] - \frac{c^2}{m^2}(1-\cos q)^2, \\
B(q) &= \frac{2k\alpha}{m}\sin q.
\end{align*}

The dispersion relation for this system is plotted in Figs.~\ref{fig:FIG7}(b) and \ref{fig:FIG7}(c). For small non-reciprocity ($\alpha \sim \mathcal{O}(\epsilon)$)  and damping ($c \sim \mathcal{O}(\epsilon)$), we approximate the dispersion relation as
\begin{align}
\text{Re}[\omega(q)] &= \sqrt{A_0(q)} + \frac{k^2 \alpha^2 \sin^2 q}{2m^2 A_0(q)^{3/2}} \nonumber \\
&\quad - \frac{c^2(1-\cos q)^2}{2m^2 \sqrt{A_0(q)}} + \mathcal{O}(\epsilon^3), \\
\text{Im}[\omega(q)] &= \frac{k\alpha \sin q}{m \sqrt{A_0(q)}} - \frac{c(1-\cos q)}{m} + \mathcal{O}(\epsilon^3). \label{eq:oddspring_onsitedamper_approx}
\end{align}
where $A_0(q) = \frac{1}{m} \left[ k_g + 2k(1 - \cos q) \right ]$. In contrast to the uniform {decrease in frequency for the onsite case in Sec.~\ref{sec:2}B}, the imaginary part of the frequency in the intersite case is {reduced} by a wavenumber-dependent term, {$c(1-\cos q)/m$}. As shown in Fig.~\ref{fig:FIG7}(c), this damping effect is zero at $q=0$ and maximal at the edge of the Brillouin zone ($q=\pm\pi$). The higher-order correction to $\text{Re}[\omega(q)]$ due to damping is again negative, similar to the onsite case. Nevertheless, $\text{Re}[\omega(q)]$ still remains a symmetric function of $q$. However, the $q$-dependent shift results in a distinct asymmetric profile for $\text{Im}[\omega(q)]$.

We show the parametric plots in Figs.~\ref{fig:FIG7}(d) and \ref{fig:FIG7}(e). While the group velocity characteristics remain largely unchanged, the boundaries separating gain and decay are qualitatively different from the onsite case. Notably, attenuation is now more pronounced near $q=\pm\pi$. Consequently, the condition for neutrally stable traveling waves ($\text{Im}[\omega] = 0$) is reshaped. By setting the imaginary part of the frequency from our approximation to zero in the long-wavelength limit ($q \to 0$), we find this condition to be
\begin{equation}
\alpha \approx \left( \frac{c}{2k} \sqrt{\frac{k_g}{m}} \right) q.
\label{eq:linear_condition}
\end{equation}
This linear relationship between $\alpha$ and $q$, in sharp contrast to the hyperbolic condition found for the onsite damper, can be verified by the boundary line near the origin in Fig.~\ref{fig:FIG7}(f).

\begin{figure}[!]
\includegraphics[width=\reprintcolumnwidth]{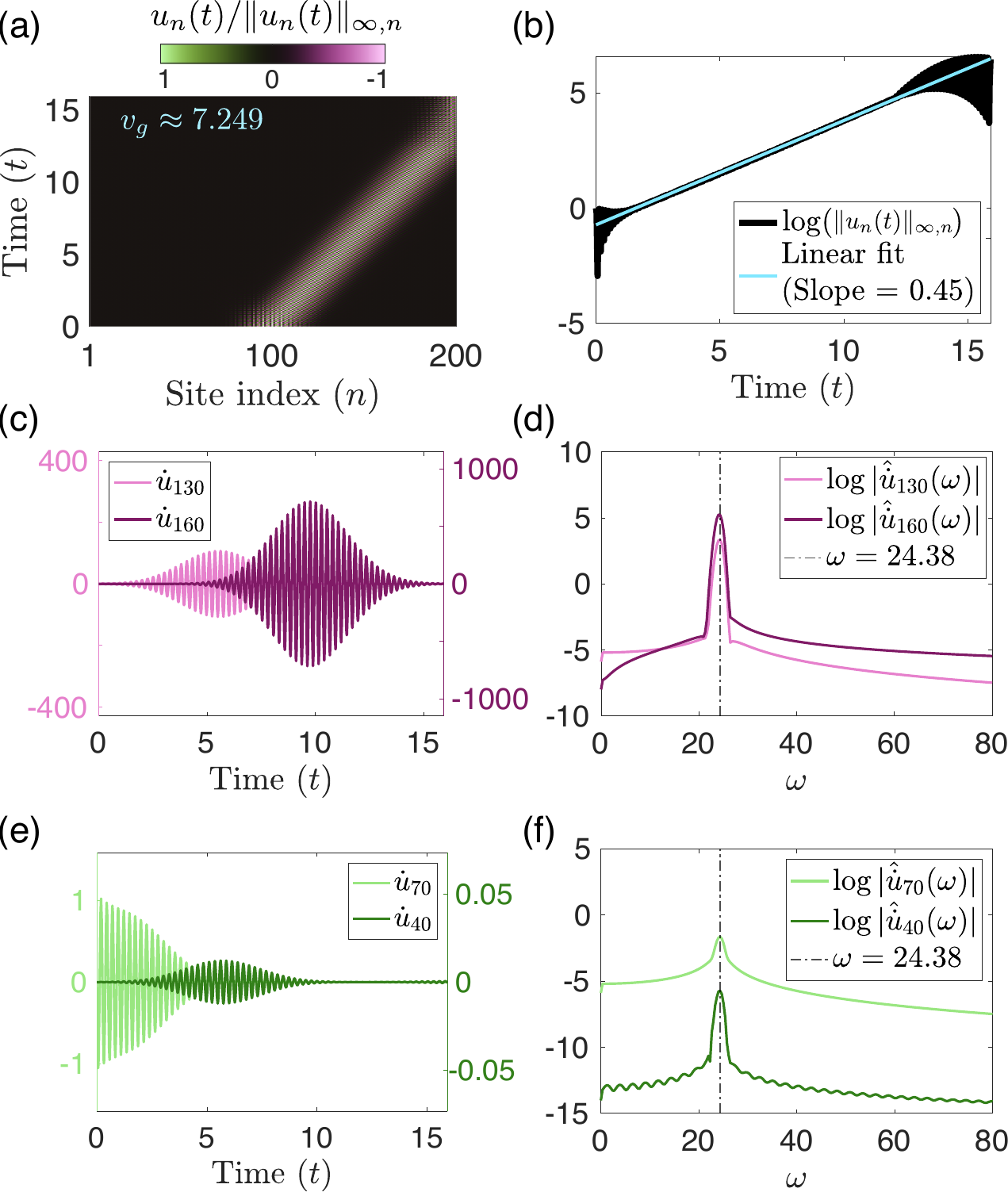}
\caption{\label{fig:FIG8}{{\textbf{Finite-chain simulations of wave propagation and directional amplification in a lattice with nonreciprocal stiffness and intersite damping.}}
(a) Space–time {evolution of the space-normalized displacement field,} showing rightward amplification of the wave packet. The measured group velocity \( v_g \approx 7.249 \) agrees with the theoretical prediction. 
(b) Logarithmic plot of the instantaneous global maximum of the displacement field, with a linear fit showing a slope of 0.45, in close agreement with the analytical growth rate. 
(c–d) Velocity profiles and corresponding spectra at downstream sites (\( n = 130, 160 \)) confirm energy amplification at the {peak} frequency \( \omega \approx 24.38 \). 
(e–f) Upstream responses at {(\( n = 70, 40 \))} show strong attenuation in both time and frequency domains.}}
\raggedright
\end{figure}

We next examine the effect of intersite damping on the wave dynamics of the finite lattice. The results, presented in Fig.~\ref{fig:FIG8}, show that unidirectional amplification of a wave packet is preserved, similar to the case with onsite damping. While the group velocity is comparable, the amplification rate differs significantly. This is consistent with the asymmetry observed in the imaginary part of the frequency, $\text{Im}(\omega)$, as shown in the dispersion diagram [Fig.~\ref{fig:FIG7}(c)]. These findings demonstrate that the nonreciprocal amplification mechanism survives in the presence of a small intersite damping, although the growth rate altered differently than for onsite damping. 


To summarize, we have investigated the role of damping in a lattice with nonreciprocal stiffness by considering three representative cases: no damping, onsite damping, and intersite damping. {The effect of onsite damping is independent of the wavenumber $q$. As seen from Eq.~\eqref{eq:damped_approx_Imag}, it contributes a constant term of $-c_g/2m$ to $\text{Im}(\omega)$, leading to uniform attenuation of all modes. In contrast, intersite damping introduces a wavenumber-dependent term of the form $-c(1-\cos q)/m$ (see Eq.~\eqref{eq:oddspring_onsitedamper_approx}). This variation produces selective attenuation across the Brillouin zone, modifying the growth and decay rates non-uniformly.}

In the next section, we shift our focus to a distinct configuration involving a reciprocal elastic lattice equipped with a nonreciprocal intersite damper. This setup reverses the roles of stiffness and damping in imparting nonreciprocity, allowing us to isolate and study the wave manipulation capabilities arising purely from dissipative asymmetry.

\section{\label{sec:3}Lattice with non-reciprocal damping}
\begin{figure}[t!]
\includegraphics[width=\reprintcolumnwidth]{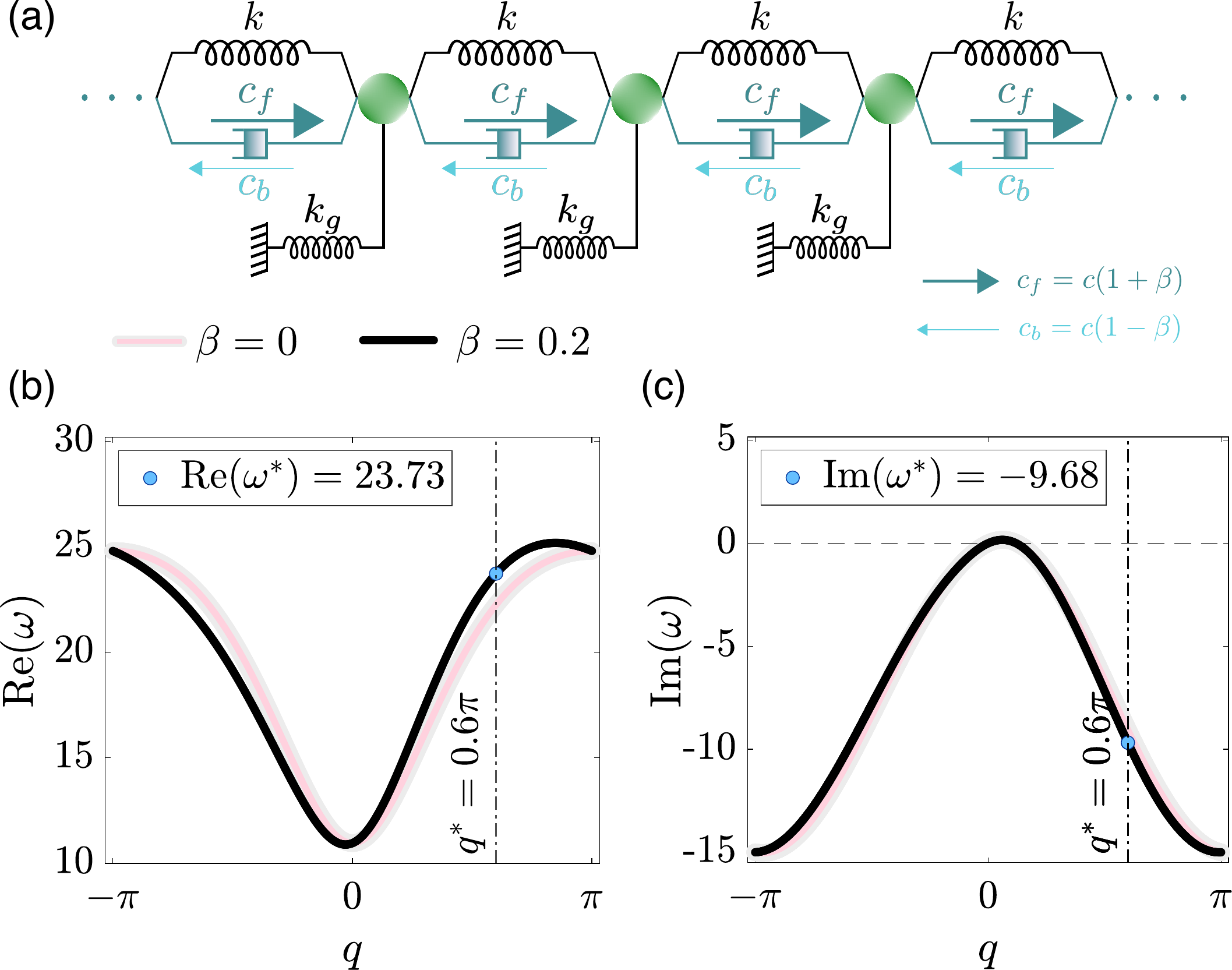}
\caption{\label{fig:responseFig2}{{\textbf{Wave propagation in a 1D lattice with nonreciprocal damping.} 
(a) Schematic of the spring–mass chain with direction-dependent damping coefficients 
\( c_f = c(1+\beta) \) and \( c_b = c(1-\beta) \), along with reciprocal intersite stiffness \( k \) and onsite stiffness \( k_g \). 
(b) Real and (c) imaginary parts of the dispersion relation \( \omega(q) \) for reciprocal (\( \beta = 0 \)) and nonreciprocal (\( \beta = 0.2 \)) cases. 
The real and imaginary parts exhibit asymmetry. The imaginary part is predominantly negative, indicating dissipation.
We take \( c = 7.5 \) and keep other parameters identical to the reciprocal case.}}}
\raggedright
\end{figure}

This section considers a system with reciprocal stiffness ($\alpha=0$) and non-reciprocal intersite damping {as shown in Fig.~\ref{fig:responseFig2}a}. The damping in the forward direction is taken to be $c_f = c(1 + \beta)$, while in the backward direction it is $c_b = c(1 - \beta)$, where $\beta \in [-1, 1]$. The governing equation of motion for this model, ignoring onsite damping, is
\begin{equation}
\begin{split}
    m \ddot{u}_{n} &+ c(2\dot{u}_{n} - (1+\beta) \dot{u}_{n-1} - (1-\beta) \dot{u}_{n+1}) \\
    &+ (2k+k_g)u_{n} -k(u_{n-1} + u_{n+1}) = 0.
\end{split}
\label{eq:EOM_nonrecip_damp}
\end{equation}
The resulting dispersion relation can be expressed as
\begin{equation}
\omega(q) = \frac{c\beta}{m}\sin q - i\frac{c(1-\cos q)}{m} + \sqrt{r(q)} e^{i \theta(q)/2}.
\end{equation}
The real and imaginary parts are given by
\begin{align}
\text{Re}[\omega(q)] &= \frac{c\beta}{m}\sin q + \sqrt{r(q)} \cos\left(\frac{\theta(q)}{2}\right), \label{eq:EQ25} \\
\text{Im}[\omega(q)] &= -\frac{c}{m}(1-\cos q) + \sqrt{r(q)} \sin\left(\frac{\theta(q)}{2}\right),
\label{eq:EQ26}
\end{align}
where
\begin{align*}
r(q) &= \sqrt{A(q)^2 + B(q)^2}, \quad \theta(q) = \operatorname{atan2}\left(B(q), A(q)\right), \\
A(q) &= \frac{1}{m}\left[k_g + 2k(1-\cos q)\right] + \frac{c^2\beta^2}{m^2}\sin^2 q \\
&\quad - \frac{c^2}{m^2}(1-\cos q)^2, \\
B(q) &= -\frac{2c^2\beta}{m^2}\sin q (1-\cos q).
\end{align*}
These results reveal a different non-reciprocal mechanism compared to the case of non-reciprocal springs. The presence of the term proportional to $\sin q$ in Eq.~\eqref{eq:EQ25} makes the oscillation frequency $\text{Re}[\omega(q)]$ also asymmetric with respect to $q=0$.
{{This spectral asymmetry implies that forward- and backward-propagating waves (at wavenumbers $\pm q$) are characterized by distinct oscillation frequencies and group velocities. Moreover, the imaginary component $\text{Im}[\omega(q)]$ also acquires an asymmetric profile for nonzero $\beta$, as illustrated in the dispersion curves in Figs.~\ref{fig:responseFig2}(b) and \ref{fig:responseFig2}(c). This indicates that the nonreciprocity parameter $\beta$ simultaneously dictates both the wave kinematics (speed/frequency) and the temporal envelope evolution (growth/decay rates).

While this general model demonstrates the broad impact of nonreciprocal damping, it highlights a limitation regarding control: $\beta$ introduces asymmetry in both the real and imaginary dispersion curves concurrently. This coupling prevents the independent manipulation of signal timing (velocity) and signal magnitude (dissipation/gain). To isolate the kinematic nonreciprocity from the dissipative effects, we next examine a specialized form of nonreciprocal damping inspired by gyroscopic mechanics.}}

\subsection*{Special non-reciprocal damping: Gyroscopic damping}
\begin{figure}[!]
\includegraphics[width=\reprintcolumnwidth]{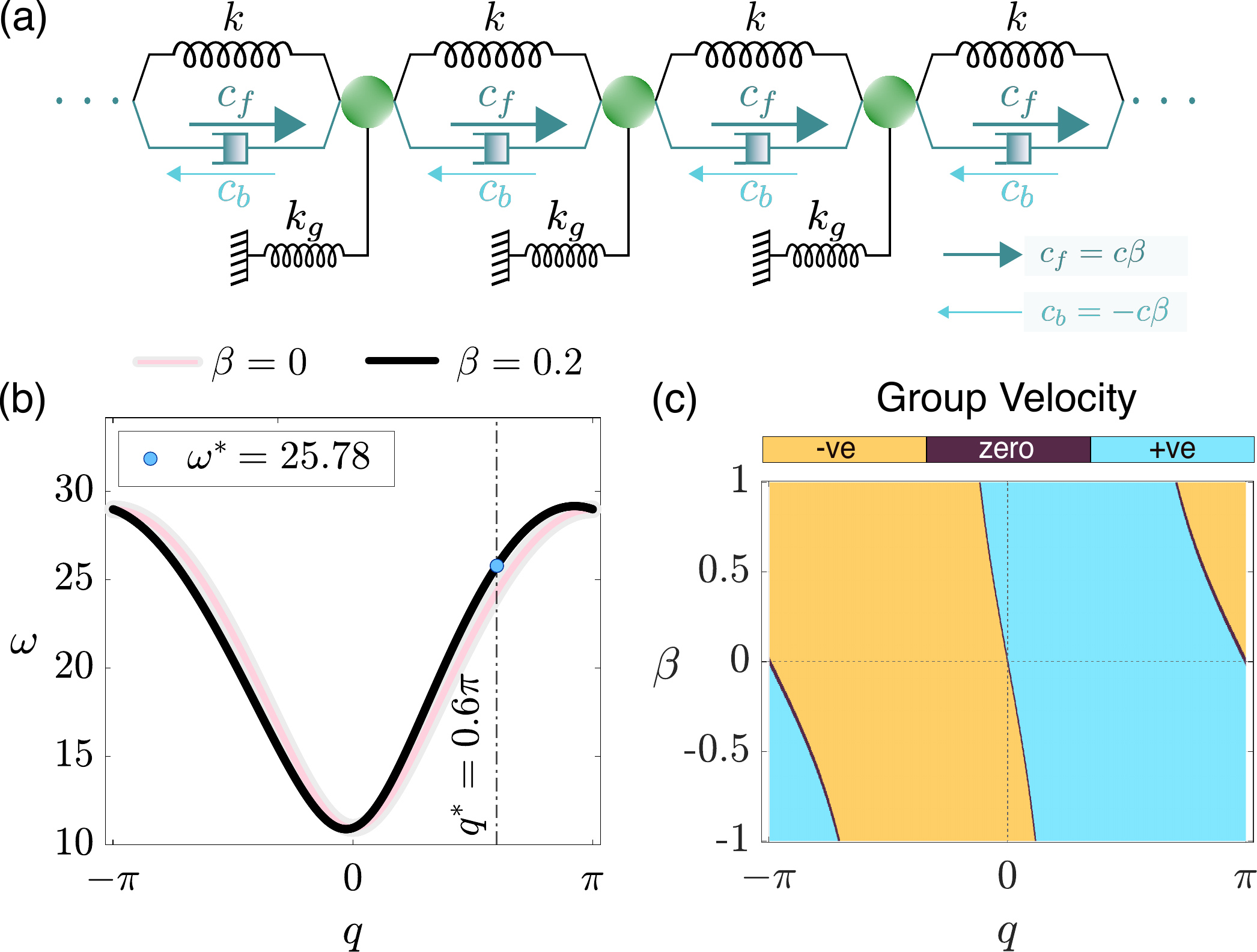}
\caption{\label{fig:FIG11}{\textbf{Wave propagation in a 1D lattice with a gyroscopic damper.} 
(a) Schematic of the lattice featuring gyroscopic damping, implemented via forward and backward dashpots with \( c_f = c\beta \) and \( c_b = -c\beta \), alongside reciprocal intersite stiffness \( k \) and onsite stiffness \( k_g \). 
(b) Purely real dispersion relation \( \omega(q) \) for \( \beta = 0 \) and \( \beta = 0.2 \). 
(c) Parametric plot showing the sign of the group velocity as a function of wavenumber \( q \) and damping asymmetry parameter \( \beta \), revealing directional bias induced by the nonreciprocal damping configuration. We take \( c = 7.5 \) and keep other parameters the same.}}
\raggedright
\end{figure}

We define a special non-reciprocal damper that leads to purely oscillatory dynamics. As depicted in Fig.~\ref{fig:FIG11}(a), this corresponds to an effective damping of $c\beta$ in the forward direction and $-c\beta$ in the backward direction. This is akin to gyroscopic damping that is not dissipative~\cite{Carta2014, Baz2020, Attarzadeh2019, Wang2015}.  This construction isolates the non-reciprocal behavior from the net dissipation inherent in standard intersite damping. In addition, this special case will later be combined with non-reciprocal stiffness to explore novel wave dynamics.

The governing equation of motion for a lattice with this gyroscopic damper is
\begin{equation}
m \ddot{u}_{n} + c\beta(\dot{u}_{n+1} - \dot{u}_{n-1}) + (2k+k_g)u_{n} - k(u_{n-1} + u_{n+1}) = 0,
\label{eq:EOM_special_damp}
\end{equation}
and the associated dispersion relation is
\begin{equation}
\omega(q) =  \frac{c\beta}{m}\sin q + \sqrt{ \frac{c^2\beta^2}{m^2}\sin^2 q + \frac{1}{m}\left[k_g + 2k(1-\cos q)\right] }.
\label{eq:disp_special_damp}
\end{equation}
The dispersion relation in Eq.~\eqref{eq:disp_special_damp} has two key features. First, the term  $\sin q$ breaks the symmetry, $\omega(q) \neq \omega(-q)$. Second, the frequency $\omega(q)$ is purely real for all $q$, indicating that the system is free of any dissipation. This frequency asymmetry is illustrated in Fig.~\ref{fig:FIG11}(b).

The group velocity consequently loses its symmetry as well. Figure~\ref{fig:FIG11}(c) shows a parametric plot demonstrating that, in contrast to the case of non-reciprocal stiffness, points of zero group velocity can emerge at wavenumbers other than $q=0$ or $q=\pm\pi$. 
{Interestingly, a similar observation is made related to the modification of Brillouin zone boundaries in a physically distinct system of a translating elastic rod~\cite{AlBa'ba'a2023}. In that work, a momentum bias introduces a velocity-dependent coupling term in the discrete framework, which is mathematically analogous to the gyroscopic damping presented in our study.}

In the limit of weak non-reciprocity ($\beta$), Eq.~\eqref{eq:disp_special_damp} can be approximated as
\begin{equation}
\omega(q) = \sqrt{A_0(q)} + \frac{c\beta}{m}\sin q + \mathcal{O}(\beta^2),
\label{eq:approx_disp}
\end{equation}
where $A_0(q) = \frac{1}{m}[k_g + 2k(1-\cos q)]$. From this approximate form, we can determine the condition for zero group velocity in the long-wavelength limit ($q \to 0$). The locus of points where $v_g(q) = 0$ is given by
\begin{equation}
\beta \approx {-} \frac{\sqrt{m} k}{c \sqrt{k_g}}q.
\label{eq:locus}
\end{equation}
This linear relationship is consistent with the region near the origin in the parametric plot shown in Fig.~\ref{fig:FIG11}(c).
\begin{figure}[t!]
\includegraphics[width=\reprintcolumnwidth]{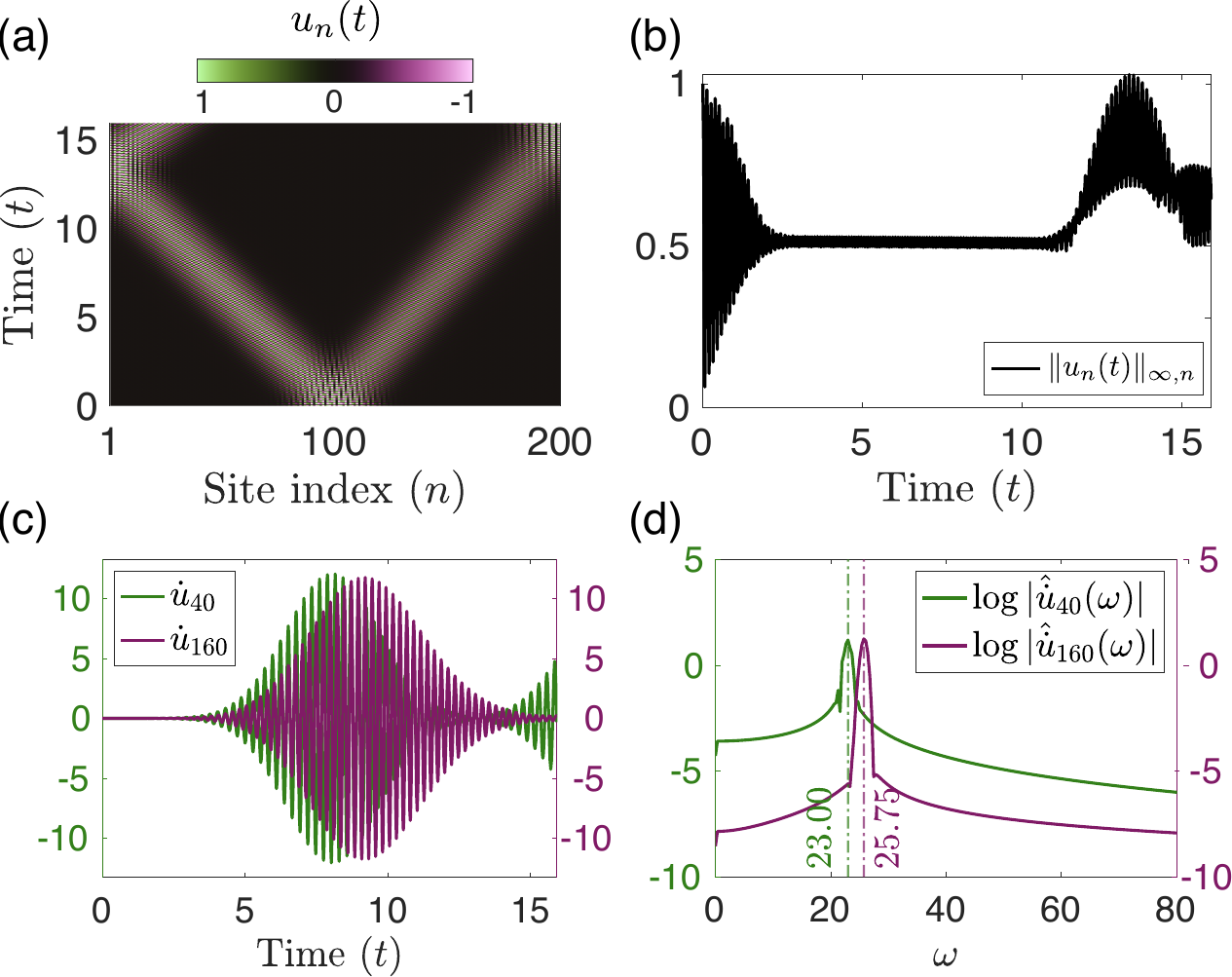}
\caption{\label{fig:FIG12}{\textbf{
{Finite-chain simulations} of wave dynamics in a 1D lattice with gyroscopic damping.} 
(a) Space-time colormap of the displacement field \( u_n(t) \) in a 200-particle finite lattice with fixed boundaries, illustrating symmetric wave amplitudes but asymmetric propagation speeds for left- and right-travelling wave packets. 
(b) Time evolution of the instantaneous global maximum, showing no net amplification or decay. 
(c–d) Velocity responses and their spectral amplitudes at sites \(n = 40\) and \(160\) reveal the coexistence of two wave components with distinct group velocities, corresponding to slightly different peak frequencies (\( \omega \approx 23.00 \) and \( \omega \approx 25.75 \)).}}
\raggedright
\end{figure}

We numerically investigate the effect of the gyroscopic damper in a finite lattice, with the results presented in Fig.~\ref{fig:FIG12}. The lattice is excited by a spatially localized Gaussian displacement near its center. The spatiotemporal evolution of the displacement field, $u_n(t)$, reveals that the initial wave packet splits into forward- and backward-propagating components [Fig.~\ref{fig:FIG12}(a)]. However, these components now propagate with different group velocities, with the backward-propagating wave traveling faster than the forward one.

As predicted by the purely real dispersion relation [Eq.~\eqref{eq:disp_special_damp}], the system is stable and exhibits no exponential growth or decay. This stability is corroborated by the global displacement norm, which remains bounded throughout the simulation [Fig.~\ref{fig:FIG12}(b)]. The velocity
 time series at two symmetric probe locations ($n=40$ and $n=160$), shown in Fig.~\ref{fig:FIG12}(c), confirms the different arrival times and shows that the wave packets maintain comparable amplitudes. A key finding is the emergence of spectral asymmetry: the velocity spectra in Fig.~\ref{fig:FIG12}(d) reveal distinct peak frequencies for the forward-propagating ($\omega \approx 25.75$) and backward-propagating ($\omega \approx 23.00$) waves. This direction-dependent {offset in frequency} is a direct consequence of the asymmetric dispersion shown in Fig.~\ref{fig:FIG11}(b).

These finite-lattice simulations, therefore, confirm that this gyroscopic damper introduces direction-dependent group velocities and {offset in frequency} without inducing exponential amplification or attenuation.

We now advance to studying a lattice system where both nonreciprocal stiffness and gyroscopic damping act simultaneously. The previous sections highlighted how each mechanism individually breaks reciprocity—nonreciprocal stiffness enabled directional amplification via complex frequencies, while the gyroscopic damper induced frequency asymmetry without introducing net gain or loss. Their combined action is expected to produce richer, more intricate wave dynamics, shaped by both directional amplification and asymmetric group velocities. In the following section, we investigate this interplay and its implications on dispersion, enhanced directional decay or growth, and wave transport in the lattice.

\section{\label{sec:4}Lattice with Combined Non-reciprocal Stiffness and Damping}
\begin{figure}[!]
\includegraphics[width=\reprintcolumnwidth]{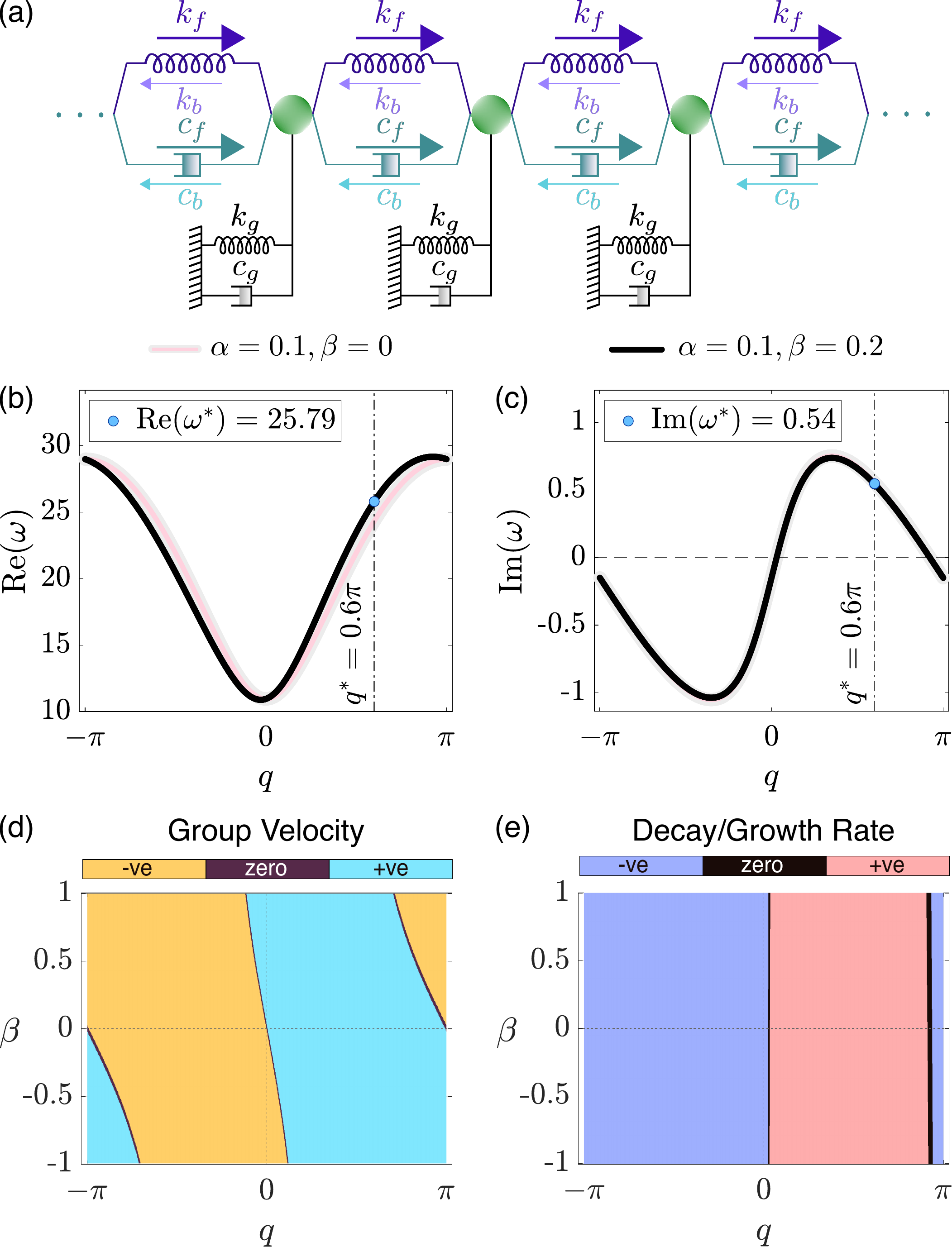}
\caption{\label{fig:FIG14}{\textbf{Wave propagation in a 1D lattice with combined nonreciprocal stiffness and damping.} 
(a) Lattice with unit cell featuring both asymmetric springs (\(k_f = k(1+\alpha), k_b = k(1-\alpha)\)) and gyroscopic dampers (\(c_f = c\beta, c_b = -c\beta\)), along with onsite stiffness \(k_g\) and damping \(c_g\). 
(b–c) Real and imaginary parts of the complex dispersion relation \(\omega(q)\) for \(\beta = 0\) and \(\beta = 0.2\) at a fixed \(\alpha = 0.1\).   
(d–e) Parametric maps of the group velocity and temporal growth/decay rate versus \(q\) and \(\beta\), demonstrating pronounced direction-dependent transport. 
We take \(c = 7.5\) and \(c_g = 0.3\). All other parameters are consistent with previous cases.}}
\raggedright
\end{figure}
We now analyze a system that combines nonreciprocal intersite stiffness (governed by \(\alpha\)) with a gyroscopic damper (governed by \(\beta\)). This configuration enables a rich interplay between directional amplification due to stiffness asymmetry and oscillation frequency asymmetry introduced by non-reciprocal damping. As illustrated in Fig.~\ref{fig:FIG14}(a), each mass in the lattice is connected to its neighbors via asymmetric springs with \(k_f = k(1+\alpha)\), \(k_b = k(1-\alpha)\), and nonreciprocal dampers with \(c_f = c\beta\), \(c_b = -c\beta\). Additionally, onsite stiffness \(k_g\) and damping \(c_g\) anchor each mass to ground. 

The governing equation of motion for the \(n\)th mass in the lattice reads:
\begin{equation}
\begin{split}
    m \ddot{u}_{n} &+ c_g\dot{u}_{n} + c\beta(\dot{u}_{n+1} - \dot{u}_{n-1}) \\
    &+ (2k+k_g)u_{n} - k(1+\alpha) u_{n-1} - k(1-\alpha) u_{n+1} = 0.
\end{split}
\label{eq:EOM_combined_final}
\end{equation}
The corresponding characteristic equation yields the exact dispersion relation:
\begin{equation}
\omega(q) = - i \frac{c_g}{2 m} + \frac{c\beta}{m}\sin q + \sqrt{r(q)} e^{i \theta(q)/2},
\end{equation}
with the real and imaginary parts given by
\begin{align}
\text{Re}[\omega(q)] &= \frac{c\beta}{m}\sin q + \sqrt{r(q)} \cos\left(\frac{\theta(q)}{2}\right), \\
\text{Im}[\omega(q)] &= - \frac{c_g}{2 m} + \sqrt{r(q)} \sin\left(\frac{\theta(q)}{2}\right),
\end{align}
where
\begin{align*}
r(q) &= \sqrt{A(q)^2 + B(q)^2}, \quad \theta(q) = \operatorname{atan2}\left(B(q), A(q)\right), \\
A(q) &= \frac{1}{m}\left[k_g + 2k(1-\cos q)\right] + \frac{c^2\beta^2}{m^2}\sin^2 q - \frac{c_g^2}{4 m^2}, \\
B(q) &= \frac{2k\alpha}{m}\sin q - \frac{c_g c \beta \sin q}{m^2}.
\end{align*}
For small values of $\alpha$, $\beta$, and $c_g$ (i.e., $\sim \mathcal{O}(\epsilon)$), a first-order Taylor expansion provides additional insight:
\begin{align}
\text{Re}[\omega(q)] &=  \sqrt{A_0(q)} + \frac{c\beta}{m}\sin q + \mathcal{O}(\epsilon^2) \label{eq:combined_Re_omega_approx_expression}\\
\text{Im}[\omega(q)] &=  \frac{k\alpha \sin q}{m \sqrt{A_0(q)}} - \frac{c_g}{2 m} + \mathcal{O}(\epsilon^2), \label{eq:combined_Im_omega_approx_expression}
\end{align}
where $A_0(q) = \frac{1}{m}[k_g + 2k(1-\cos q)]$.

These results reveal a remarkable decoupling of the two non-reciprocal effects at leading order. The imaginary part of the frequency, which governs temporal growth or decay, is determined entirely by the non-reciprocal stiffness parameter \(\alpha\), and matches the expression for lattice with nonreciprocal stiffness and onsite damping. The real part, on the other hand, acquires an asymmetric contribution from the non-reciprocal damper \(\beta\), inducing a direction-dependent {offset in frequency} without altering the system’s gain or loss characteristics. This enables independent tuning of wave amplification and oscillation frequency through parameters \(\alpha\) and \(\beta\), respectively.

Figures~\ref{fig:FIG14}(b) and \ref{fig:FIG14}(c) present the real and imaginary parts of the complex dispersion relation, $\omega(q)$, for a fixed stiffness asymmetry ($\alpha=0.1$) to investigate the influence of the parameter $\beta$. While the imaginary part, $\text{Im}[\omega(q)]$, is nearly identical to the case where $\beta=0$, the real part, $\text{Re}[\omega(q)]$, becomes asymmetric for non-zero values of $\beta$.

This behavior is further detailed in the parametric maps shown in Figs.~\ref{fig:FIG14}(d) and \ref{fig:FIG14}(e). The group velocity map [Fig.~\ref{fig:FIG14}(d)] exhibits a strong asymmetry in both sign and magnitude that is tunable by $\beta$. In contrast, the growth rate map [Fig.~\ref{fig:FIG14}(e)] confirms that $\text{Im}[\omega(q)]$ is largely insensitive to $\beta$. The inherent asymmetry in the growth rate with respect to the wavenumber $q$ is therefore preserved, as it is governed by the nonreciprocal stiffness $\alpha$.

\begin{figure}[t!]
\includegraphics[width=\reprintcolumnwidth]{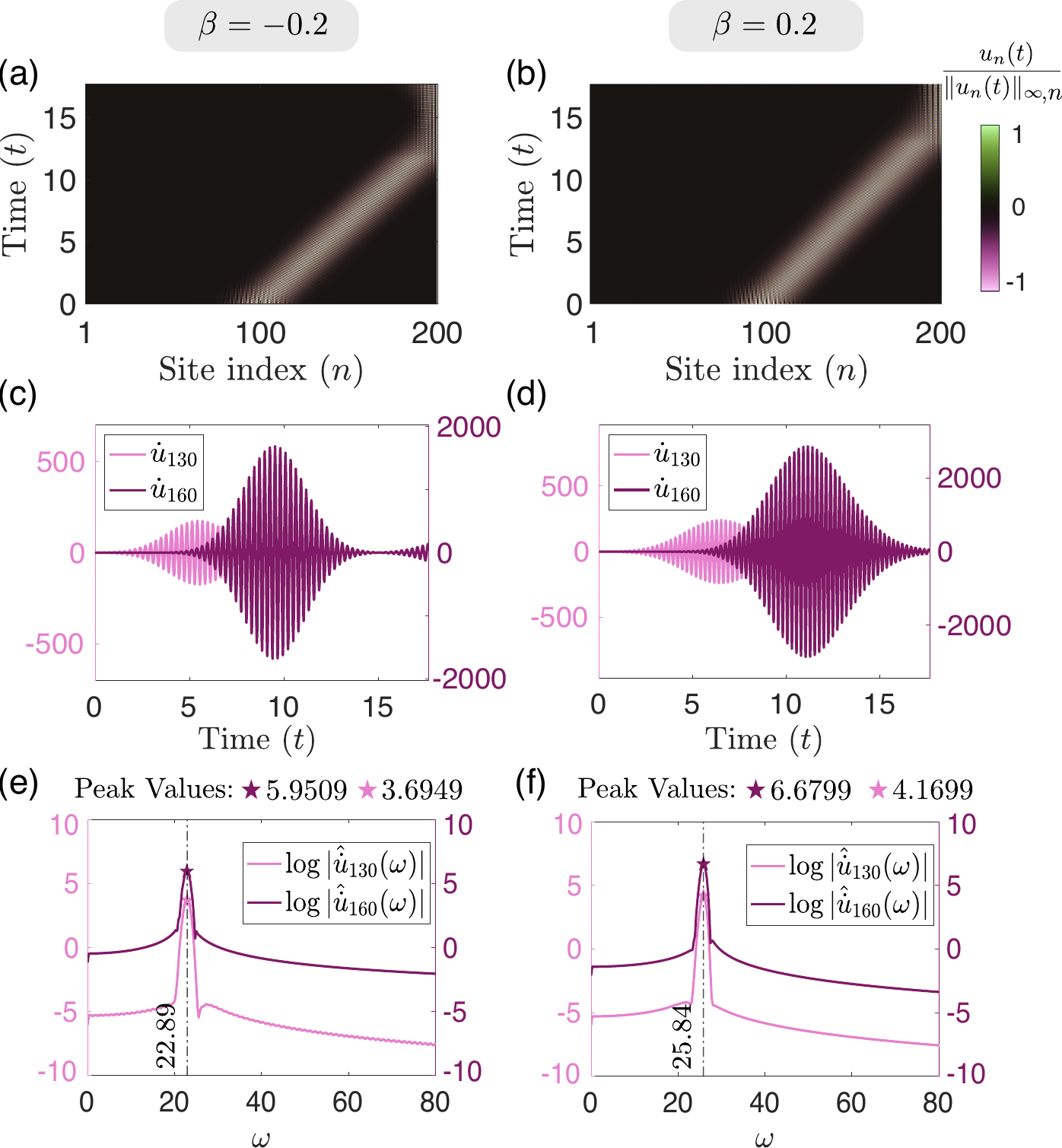}
\caption{\label{fig:FIG15}{\textbf{Finite lattice simulation illustrating amplification dependence on $\beta$ in a system with combined nonreciprocal stiffness and gyroscopic damping.}  
(a,b) Space–time plots of normalized displacement for \(\beta = \pm 0.2\). 
(c,d) Corresponding time-domain velocity responses at \(n = 130, 160\). The case with $\beta=0.2$ shows a slightly more amplified wavepacket due to slower group velocity. 
(e,f) Velocity spectra reveal {peak} frequencies. We take \(\alpha = 0.1\) and keep other parameters the same as before.
}}
\raggedright
\end{figure}

We present finite chain simulations in Fig.~\ref{fig:FIG15} to investigate the interplay between nonreciprocal stiffness and nonreciprocal damping. The simulations are performed for two contrasting cases of the damping parameter, $\beta = -0.2$ and $\beta = +0.2$, while the stiffness asymmetry is held constant at $\alpha = 0.1$. As in previous sections, the system is excited by a Gaussian-modulated initial displacement at its center.

The results reveal that the sign of $\beta$ provides a mechanism to control wave dynamics. Figures~\ref{fig:FIG15}(a) and \ref{fig:FIG15}(b) show that the forward-propagating wave is amplified in both cases, a characteristic governed by the nonreciprocal stiffness $\alpha$. However, switching the sign of $\beta$ alters the group velocity of the wave packet. This is explicitly confirmed by the velocity time series at downstream sites ($n=130$ and $n=160$) in Figs.~\ref{fig:FIG15}(c) and (d), which show that the wave packet for $\beta=-0.2$ arrives at a given site faster than for $\beta=+0.2$. 
{This is consistent with the evolution of temporal peak for the forward propagating amplified wave packet (see Fig.~\ref{fig:FIGA1} in Appendix A).}

Interestingly, while the amplification rate is set by $\alpha$, the net amplification at a specific site depends on the group velocity as well. For $\beta=+0.2$, the wave packet travels slower, allowing more time for the amplification, resulting in a larger amplitude at a given location compared to the $\beta=-0.2$ case. The net amplification {(seen from the temporal peaks of the Gaussian packets)} is therefore a combined effect of the growth rate (dictated by $\alpha$) and the propagation time (dictated by $\beta$). This is one of the key results of our work.

\begin{figure}[t!]
\includegraphics[width=\reprintcolumnwidth]{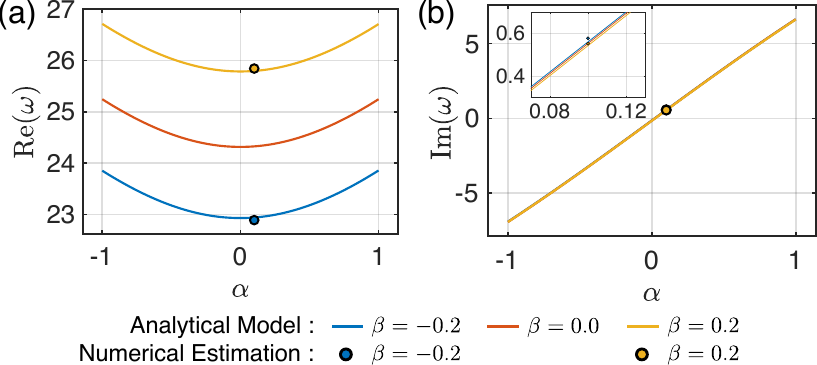}
\caption{\label{fig:FIG16}{\textbf{Effect of $\alpha$ and $\beta$ on oscillation frequency and temporal growth at \( q = 0.6\pi \).}  
(a) \(\text{Re}(\omega)\) plotted as a function of stiffness asymmetry \(\alpha\) for \(\beta = -0.2, 0, 0.2\). Change in $\beta$ leads to a significant change in oscillation frequency for a fixed $\alpha$.  
(b) \(\text{Im}(\omega)\) for the same parameter sweep shows minimal variation, suggesting that damping asymmetry has a negligible effect on the temporal growth rate.  
Solid lines represent analytical results from the complex dispersion relation, and circular markers indicate numerically extracted peak frequencies from finite-lattice simulations (see Fig.~\ref{fig:FIG15}).  
The inset in (b) confirms close agreement between analytical and numerical growth rates at \(\alpha = 0.1\).}}
\raggedright
\end{figure}

Finally, Figs.~\ref{fig:FIG15}(e) and \ref{fig:FIG15}(f) show that the sign of $\beta$ also shifts the {peak} frequency of the amplified wave, along with a gain in the spectral content.
 This tunability is further explored in Fig.~\ref{fig:FIG16}, which plots the amplified {peak} frequency as a function of $\alpha$ and $\beta$ for a forward-propagating wave. The frequency exhibits a nearly quadratic dependence on $\alpha$ for a fixed $\beta$ [Fig.~\ref{fig:FIG16}(a)], consistent with the higher-order corrections in Eq.~\eqref{eq:damped_approx_Real}. In contrast, Fig.~\ref{fig:FIG16}(b) demonstrates that varying $\beta$ provides an effective route to tune the wave's frequency while leaving the amplification rate, governed by $\alpha$, nearly unchanged.

In summary, while the amplification rate of waves is primarily determined by $\alpha$, the sign and magnitude of $\beta$ can be used to tune net amplification (through changing the group velocity) and the {peak  } frequency of the wave packet. 

\begin{figure*}[t]
\includegraphics[width=\textwidth]{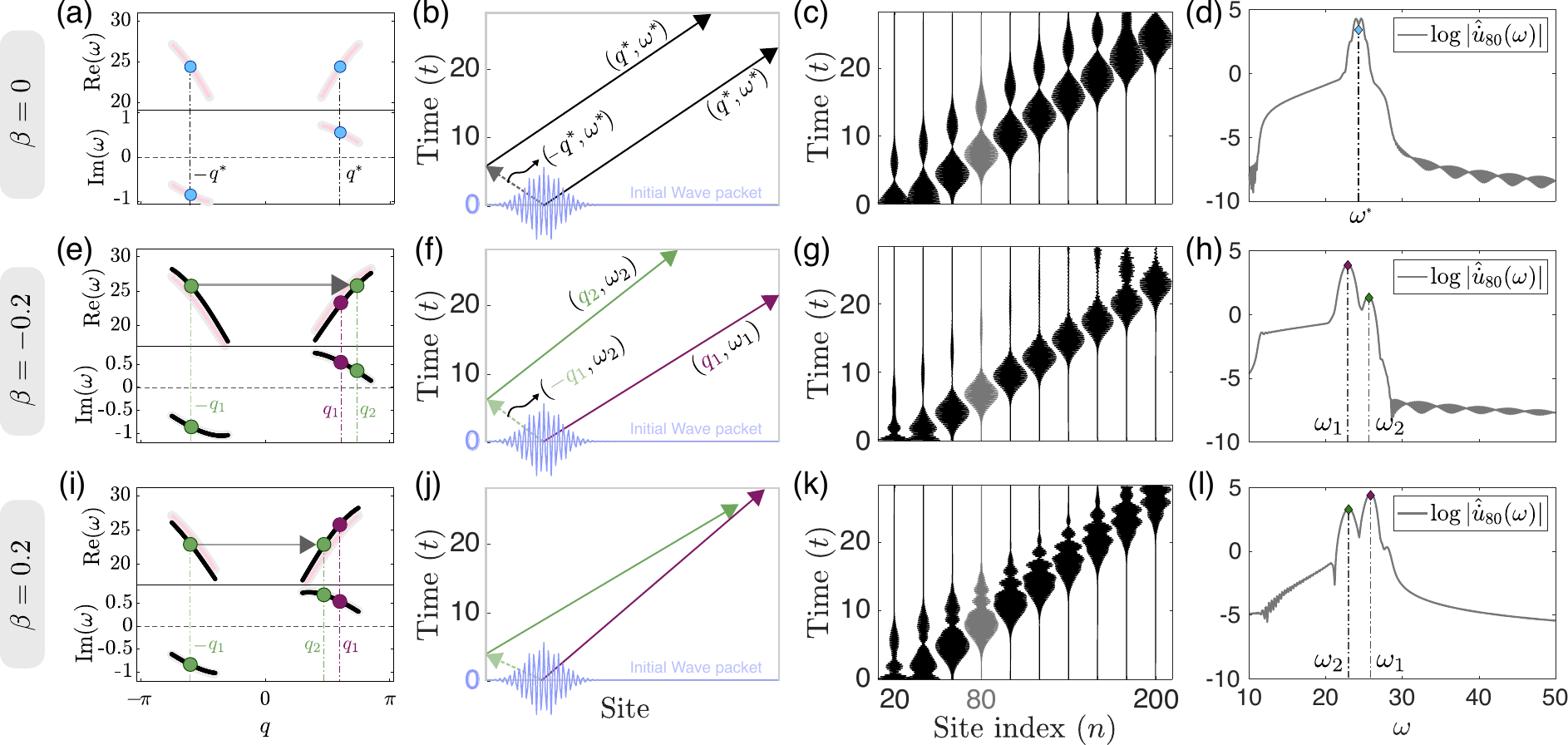}
\caption{\label{fig:FIG18}{\textbf{Boundary-induced interference of forward and backward wave packets at $|q^*|=0.6\pi$ in a 1D lattice with combined nonreciprocal stiffness and gyroscopic damping.}
(a,e,i) Complex dispersion relations identifying the primary forward ($+q^*$) and backward ($-q^*$) components excited by the initial Gaussian packet, alongside the secondary wavenumber $q_2$ generated upon boundary reflection.
(b,f,j) Ray diagrams illustrating the kinematic trajectories of the direct and reflected propagation paths.
(c,g,k) Time-normalized velocity evolution, illustrating the asymmetric amplification and the resulting interference between incident and reflected packets.
(d,h,l) Velocity spectra measured at site $n=80$, displaying single or dual frequency peaks depending on the damping parameter $\beta$.
Key spectral peaks:
$\beta=0$: $(q^*,\omega^*)=(0.6\pi, 24.38)$;
$\beta=-0.2$: primary $q_1=0.6\pi$ ($22.89$) and reflected $q_2=0.735\pi$ ($25.63$);
$\beta=+0.2$: primary $q_1=0.6\pi$ ($25.86$) and reflected $q_2=0.481\pi$ ($23.00$).}}
\raggedright
\end{figure*}

\subsection*{Wave {interference} through boundary reflection}
In a finite system, boundaries are not merely passive reflectors but can be actively utilized to manipulate wave properties. Here, we investigate how boundary reflection can induce wave {interference}, providing an additional mechanism for tuning the amplified signal's frequency content. {{In this context, interference arises from the superposition of the primary forward-propagating wave and the reflected backward-propagating wave, which acquire distinct oscillation frequencies due to the asymmetric dispersion.}} We achieve this by launching a wave packet near one end of the lattice. The backward-propagating (attenuated) component travels to the opposite boundary, reflects, and then propagates forward, {interfering} with the original forward-propagating (amplified) component. As we will show, the nonreciprocal damping parameter $\beta$ is crucial in controlling this {interference}. {{Figure~\ref{fig:FIG18} compares three representative cases: $\beta = 0$ (top row), $\beta = -0.2$ (middle row), and $\beta = +0.2$ (bottom row), while keeping the stiffness asymmetry fixed at $\alpha = 0.1$.

For the case without gyroscopic damping ($\beta=0$), the dynamics are governed solely by the stiffness asymmetry $\alpha$. The dispersion and ray diagrams [Figs.~\ref{fig:FIG18}(a--b)] show that the forward and backward components originate from wavenumbers $\pm q^*$. Upon reflection, the backward pulse conserves its frequency but scatters into a forward mode with wavenumber $+q^*$. Consequently, the reflected packet inherits the exact group velocity and growth rate of the primary forward packet. This identical amplification is evident in the time-normalized waterfall plot [Fig.~\ref{fig:FIG18}(c)], where the relative amplitudes of the two pulses (forward and reflected) remain constant. Accordingly, the velocity spectrum [Fig.~\ref{fig:FIG18}(d)] at the probe site $n=80$ exhibits a single peak centered around $\omega^*$.

For $\beta=-0.2$, the initial wave packet splits into components at $+q_1$ and $-q_1$ with distinct frequencies. As shown in Fig.~\ref{fig:FIG18}(e), the backward pulse oscillates at a higher frequency than the forward one ($\mathrm{Re}[\omega(-q_1)] > \mathrm{Re}[\omega(+q_1)]$). After reflecting at the boundary, this high-frequency component couples to a forward mode with positive wavenumber $q_2$. Crucially, this mode ($q_2$) corresponds to a slower group velocity and a lower growth rate compared to the primary forward packet ($q_1$). The ray diagram [Fig.~\ref{fig:FIG18}(f)] illustrates the resulting divergent trajectories, which prevent the reflected packet from overlapping with the primary pulse. This is confirmed by the waterfall plot [Fig.~\ref{fig:FIG18}(g)], where the reflected wave lags behind and its relative amplitude decays. Consequently, the spectrum at $n=80$ [Fig.~\ref{fig:FIG18}(h)] displays two peaks: a dominant one at $\omega_1$ (primary wave) and a minor one at $\omega_2$ (reflected wave).

Finally, for $\beta=+0.2$, the initial wave packet generates forward and backward components at $+q_{1}$ and $-q_{1}$, but the dispersion diagram now indicates that the backward pulse oscillates at a lower frequency ($\mathrm{Re}[\omega(-q_{1})] < \mathrm{Re}[\omega(+q_{1})]$) [Fig.~\ref{fig:FIG18}(i)]. Upon boundary reflection, this lower-frequency pulse scatters into a forward mode at wavenumber $q_{2}$, which possesses a higher group velocity and a larger temporal growth rate than the primary forward wave at $q_{1}$. The ray diagram [Fig.~\ref{fig:FIG18}(j)] reveals converging trajectories, where the faster reflected packet catches up to the slower primary packet, facilitating spatial overlap. Because the reflected mode grows faster than the primary forward mode, its relative amplitude strengthens more rapidly as the two packets approach each other. This combined effect of convergence and differential growth is directly visible in the waterfall plot [Fig.~\ref{fig:FIG18}(k)], where the reflected component becomes increasingly dominant in the overlap region. The resulting FFT at $n=80$ [Fig.~\ref{fig:FIG18}(l)] exhibits two prominent peaks at $\omega_1$ and $\omega_2$. Notably, the reflected-mode peak ($\omega_2$) is substantially stronger here than in the $\beta=-0.2$ case due to its enhanced growth rate. Thus, the combination of boundary reflection and damping asymmetry produces a robust, tunable two-frequency interference response.}}

The value of $\beta$ along with the boundary thus acts as a switch, controlling which frequencies are mixed and their relative dominance. Therefore, these results demonstrate that boundaries in nonreciprocal lattices can be seen as functional elements that enable complex wave-shaping phenomena in addition to wave amplification.

\section{\label{sec:5}Conclusion}

This work presents a comprehensive framework for controlling wave propagation in one-dimensional lattices by combining nonreciprocal stiffness and engineered damping. We first established that conventional onsite and intersite viscous damping mechanisms, while modifying the system's dispersion, primarily counteract the gain from nonreciprocal stiffness, thereby offering limited control over wave propagation. In contrast, by introducing a non-dissipative gyroscopic damping mechanism, we uncovered our central result: a robust decoupling of control. This allows the nonreciprocal stiffness ($\alpha$) to dictate the temporal growth rate, while the gyroscopic damping ($\beta$) independently tunes the wave's group velocity and oscillation frequency.

This decoupling of temporal gain from wave kinematics enables novel phenomena. By tuning $\beta$, the travel time of a wave packet through the amplifying medium can be controlled, leading to the key insight that slower-propagating waves accumulate greater net amplification due to prolonged exposure to the gain mechanism. Furthermore, we demonstrated that boundaries in finite systems can be functionalized to induce wave {interference through divergent and convergent reflected waves. Specifically, the parameter $\beta$ governs the efficacy of this interference by acting as a switch to generate either diverging reflected paths (preventing interference) or converging paths (facilitating overlap) with varying growth rates, thereby enabling the selective mixing and amplification of distinct frequency components.} These behaviors, predicted by our analytical models, were consistently validated by time-domain simulations.

These findings provide a versatile toolkit for programming wave behavior, with significant implications for the design of active metamaterials. The ability to independently manage gain, group velocity, and frequency offers a new paradigm for engineering systems with on-demand functionalities, including advanced unidirectional amplifiers, and reconfigurable {acoustic and vibration isolators}.

Future research could explore experimental realization of this model, potentially through active electronic circuits or feedback-controlled mechanical systems. A compelling direction exploration of topological features and bulk-boundary correspondence in these lattices and their extension to higher dimensions. Finally, investigating the influence of disorder, nonlocality~\citep{rosa2020dynamics}, and nonlinearity~\cite{Veenstra2024} on these systems remains a rich area for exploration, promising to uncover more complex spatiotemporal wave dynamics.

\begin{acknowledgments}
{We thank Prof. Nicholas Boechler and Prof. Nicholas Gravish (University of California, San Diego) and Dr. Sayan Jana (Tel Aviv University) for insightful discussions.} R.C. gratefully acknowledges support from the Indian Institute of Science Startup Grant.
\end{acknowledgments}

\section*{Author Declaration}
{\subsection*{CRediT authorship contribution statement}
\textbf{Harshit Kumar Sandhu}: Conceptualization, Methodology, Software, Formal Analysis, Visualization, Writing -- Original Draft, Review \& Editing;
\textbf{Saurav Dutta}: Methodology (Supporting), Software (Supporting); 
\textbf{Rajesh Chaunsali}: Conceptualization, Formal Analysis, Supervision, Writing -- Review \& Editing, Funding acquisition.}

\subsection*{Conflict of Interest} 
The authors have no conflicts to disclose.

\section*{DATA AVAILABILITY}
The data that support the findings of this study are available from the authors upon reasonable request.

\appendix
\section{Wave-packet evolution and definitions of $n_{\text{peak}}(t)$ and $t_{\text{peak}}(n)$}

We consider wave evolution in a one-dimensional discrete lattice whose dynamics are governed by a complex dispersion relation, as introduced in the main text. When a wave packet propagates through such a system, each of its Fourier components not only propagates according to its phase velocity but also undergoes spectral amplification or attenuation depending on its wavenumber. Consequently, the traveling wave packet requires careful characterization. In particular, the apparent motion of the packet can be interpreted either through the propagation of its spatial maximum, $n_{\text{peak}}(t)$, or through the arrival time of its temporal maximum at a fixed site, $t_{\text{peak}}(n)$. The distinction between the trajectories of these two parameters in the spatiotemporal map characterizes the non-Hermitian nature of the system compared to its Hermitian counterparts.

To quantitatively distinguish these behaviors, we utilize the discrete Fourier representation. We define the complex displacement field (analytic signal) $u_n(t)$, such that the physical displacement is $\mathrm{Re}[u_n(t)]$. The time evolution of a packet centered initially at $n_0$ in a lattice of $N$ particles is given by
\begin{equation}
u_n(t) = \sum_{m=0}^{N-1} \tilde{u}(q_m)\, e^{\,i(q_m (n-n_0) - \omega(q_m)t)},
\label{eq:A_time_evolution}
\end{equation}
where $\tilde{u}(q_m)$ is the spectral amplitude, and the admissible wavenumbers are $q_m = 2\pi m/N$ for $m = 0,1,\dots,N-1$.

We consider a packet with a narrow bandwidth centered at a carrier wavenumber $q^*$. We can write the spectral amplitude as
\begin{equation}
\tilde{u}(q_m) = \tilde{A}(q_m - q^*) \equiv \tilde{A}(k_m),
\end{equation}
where $k_m = q_m - q^*$. We expand the dispersion relation $\omega(q^* + k_m)$ in a Taylor series around $k_m = 0$, truncating at the first order:
\begin{equation}
\omega(q^* + k_m) \approx \omega(q^*) + \frac{d\omega}{dq}\bigg|_{q^*} k_m.
\label{eq:A_taylor}
\end{equation}
This approximation is valid provided that higher-order terms, such as group-velocity dispersion ($\omega''$), remain negligible over the propagation timescale.

In the case of a complex dispersion relation, we define the components explicitly as:
\begin{align}
\omega(q^*) &= \omega_R(q^*) + i\,\omega_I(q^*), \\
\frac{d\omega}{dq}\bigg|_{q^*} &= \omega'_R(q^*) + i\,\omega'_I(q^*).
\end{align}
Here, $\omega_R(q^*)$ and $\omega_I(q^*)$ denote the oscillation frequency and the temporal growth (or decay) rate, respectively. Similarly, $\omega'_R(q^*)$ represents the standard (real) group velocity, while $\omega'_I(q^*)$ is its imaginary counterpart.

Substituting the expansion \eqref{eq:A_taylor} into Eq.~\eqref{eq:A_time_evolution} yields:
\begin{align}
u_n(t) &\approx \sum_{m=0}^{N-1} \tilde{A}(k_m)\,
e^{\,i[(q^* + k_m)(n-n_0) - (\omega(q^*) + \omega'(q^*)k_m)t]} \nonumber \\
&= e^{\,i[q^* (n-n_0) - \omega(q^*)t]}
\sum_{m=0}^{N-1} \tilde{A}(k_m)\,
e^{\,i k_m [(n-n_0) - \omega'(q^*) t]}.
\label{eq:A9}
\end{align}

The evolution of the wave packet is governed by two distinct factors. First, the global phase factor describes the carrier wave:
\begin{equation}
e^{\,i(q^* (n-n_0) - \omega(q^*)t)}
= \underbrace{e^{\,\omega_I(q^*) t}}_{\text{Growth/Decay}}
  \cdot
  \underbrace{e^{\,i(q^* (n-n_0) - \omega_R(q^*)t)}}_{\text{Oscillation}}.
\label{eq:A10}
\end{equation}
This term dictates the propagation of the fast-oscillating carrier with phase velocity $v_p = \omega_R(q^*)/q^*$ and a uniform exponential growth or decay of the entire packet.

Second, the summation term governs the evolution of the packet's envelope:
\begin{equation}
I(n,t) = \sum_{m=0}^{N-1} \tilde{A}(k_m)\, e^{E(k_m)}, \label{eq:A11}
\end{equation}
where the exponent $E(k_m)$ is given by
\begin{equation}
E(k_m) = i k_m \big((n-n_0) - \omega'(q^*) t\big).
\label{eq:A12}
\end{equation}
Separating the real and imaginary parts of the group velocity $\omega'$, we obtain:
\begin{equation}
E(k_m)
= \underbrace{i k_m\big((n-n_0) - \omega'_R(q^*) t\big)}_{\text{Envelope Translation}}
+ \underbrace{k_m\,\omega'_I(q^*) t}_{\text{Spectral Skewness}}.
\label{eq:A13}
\end{equation}
The first term is responsible for the rigid translation of the envelope at the real group velocity $\omega'_R(q^*)$. The second term, being real, acts as a $k_m$-dependent amplification factor. For $\omega'_I > 0$, spectral components with $k_m > 0$ are amplified relative to $k_m < 0$, causing a modification of the packet shape or a shift in the effective carrier wavenumber (spectral skewness).

\subsection*{Gaussian modulated pulse}
We now analyze the specific case used in the main text: a Gaussian-modulated cosine wave. The initial physical displacement is
\begin{equation}
u_n(0) = 2C\,\exp\!\left[-\frac{(n-n_0)^2}{2\sigma_n^2}\right] \cos(q_0 (n-n_0)).
\label{eq:A_initial_gaussian}
\end{equation}
This can be decomposed into a superposition of right- ($R$) and left- ($L$) moving complex packets, $u_n(t) = u^{(R)}_n(t) + u^{(L)}_n(t)$. Focusing on the right-moving component centered at $q^* = q_0$:
\begin{equation}
u^{(R)}_n(0) = C\,\exp\!\left[-\frac{(n-n_0)^2}{2\sigma_n^2}\right] e^{\,i q_0 (n-n_0)}.
\end{equation}
The corresponding spectral amplitude $\tilde{A}(k_m)$ is Gaussian:
\begin{equation}
\tilde{A}(k_m) = \tilde{C}\,e^{-k_m^2/(2\sigma_k^2)},
\end{equation}
where the spectral width satisfies $\sigma_k = 1/\sigma_n$.
Substituting this into the envelope sum $I_R(n,t)$ and converting the summation to an integral (valid in the continuum limit for large $N$), we evaluate:
\begin{equation}
I_R(n,t) \approx \int_{-\infty}^{\infty} e^{-\frac{k_m^2}{2\sigma_k^2}}\, e^{\,i k_m ((n-n_0) - \omega'_R t) + k_m \omega'_I t} \, dk_m.
\end{equation}
Using the standard Gaussian integral solution $\int e^{-ax^2+bx} dx = \sqrt{\pi/a}e^{b^2/4a}$, and retaining only the magnitude terms, we find:
\begin{align}
|u^{(R)}_n(t)|
&\propto e^{\,\omega_I t} \cdot \exp\!\left[\frac{\sigma_k^2}{2} \left((\omega'_I t)^2 - \big((n-n_0) - \omega'_R t\big)^2 \right) \right].
\end{align}
Substituting $\sigma_k = 1/\sigma_n$ leads to the final expression for the magnitude evolution:
\begin{equation}
|u^{(R)}_n(t)| \propto\;
\exp\!\left(
\omega_I t + \frac{(\omega'_I t)^2}{2\sigma_n^2}
\right)
\times
\exp\!\left(
-\frac{\big[(n-n_0) - \omega'_R t\big]^2}{2\sigma_n^2}
\right).
\label{eq:A24}
\end{equation}
Equation~\eqref{eq:A24} reveals that within the first-order approximation, the packet retains its Gaussian spatial profile with constant width $\sigma_n$. However, the amplitude is modulated by both the standard growth rate $\omega_I$ and a correction term dependent on $\omega'_I$.

\subsection*{Two measures of packet peaks}
We now highlight the distinction between two different definitions of the wave packet's position.

\paragraph*{i. Spatial peak (at fixed $t$).}
At a fixed instant $t$, the spatial center of the envelope, $n_{\text{peak}}$, is found by maximizing $|u^{(R)}_n(t)|$ with respect to $n$. This corresponds to maximizing the spatial Gaussian term in Eq.~\eqref{eq:A24}, which occurs when the argument vanishes:
\begin{equation}
(n-n_0) - \omega'_R(q_0) t = 0
\quad\Rightarrow\quad
n_{\text{peak}}(t) = n_0  + \omega'_R(q_0)\,t.
\label{eq:A25}
\end{equation}
This confirms that the spatial envelope propagates at the real group velocity $v_g = \omega'_R(q_0)$, unaffected by the gain/loss mechanisms.

\paragraph*{ii. Temporal peak (at fixed $n$).}
Alternatively, an observer at a fixed lattice site $n$ records the time history $|u^{(R)}_n(t)|$ and identifies the time $t_{\text{peak}}$ when the signal is maximal. Maximizing the logarithm of the magnitude $G(t) = \ln |u^{(R)}_n(t)|$:
\begin{equation}
G(t) = \omega_I t
      + \frac{(\omega'_I)^2}{2\sigma_n^2} t^2
      - \frac{((n-n_0) - \omega'_R t)^2}{2\sigma_n^2}.
\end{equation}
Setting $dG/dt = 0$ yields the condition:
\begin{equation}
\omega_I + \frac{(\omega'_I)^2}{\sigma_n^2} t_{\text{peak}} - \frac{((n-n_0) - \omega'_R t_{\text{peak}})(-\omega'_R)}{\sigma_n^2} = 0.
\end{equation}
Multiplying by $\sigma_n^2$ and solving for $t_{\text{peak}}$:
\begin{equation}
t_{\text{peak}}(n)
= \frac{\omega'_R (n-n_0) + \omega_I \sigma_n^2}
       {(\omega'_R)^2 - (\omega'_I)^2}.
       \label{eq:A29}
\end{equation}
This expression differs significantly from the standard definition. Only in the conservative limit ($\omega_I = 0, \omega'_I = 0$) do we recover the familiar relation $t_{\text{peak}} = (n-n_0)/\omega'_R$.

Furthermore, assuming the imaginary component of the group velocity is small ($\omega'_I \ll \omega'_R$), Eq.~\eqref{eq:A29} can be approximated as:
\begin{equation}
t_{\text{peak}}(n) \approx \frac{n-n_0}{\omega'_R} + \frac{\omega_I \sigma_n^2}{(\omega'_R)^2}.
\label{eq:A30}
\end{equation}
This result implies that the temporal peak travels with the standard inverse group velocity (the slope $dt/dn$), but its arrival time is shifted by an intercept term $t_{\text{int}} = \frac{\omega_I \sigma_n^2}{(\omega'_R)^2}$. This delay (or advance) is determined by the interplay between the temporal growth rate ($\omega_I$), real group velocity ($\omega'_{R}$) and the packet width ($\sigma_n$), a signature of non-Hermitian wave dynamics.



\begin{figure}[h!]
\includegraphics[width=\reprintcolumnwidth]{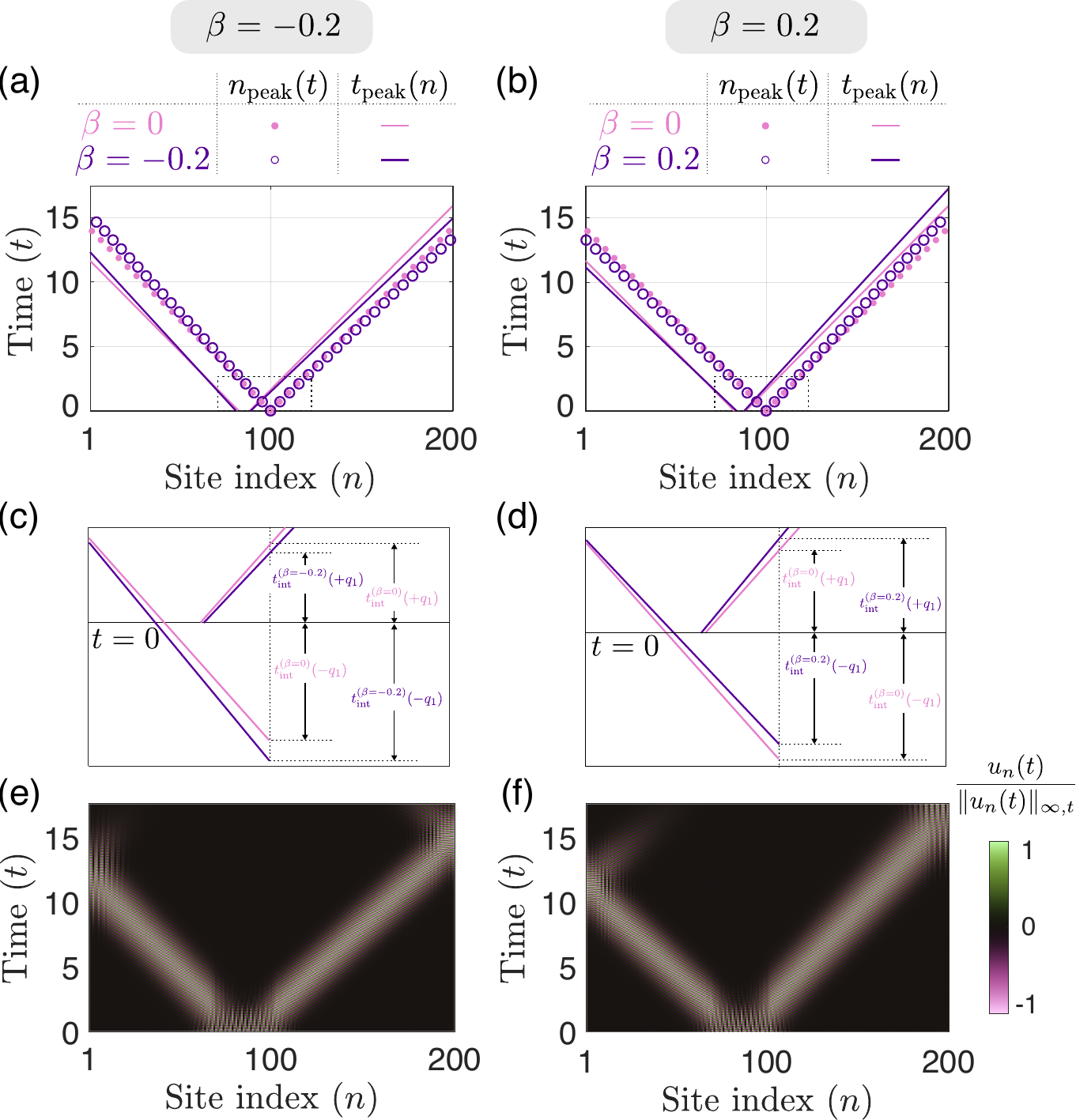}
\caption{\label{fig:FIGA1}{{\textbf{Analytical and numerical verification of peak trajectories in a 1D lattice with combined nonreciprocity.}
(a,b) Analytical trajectories of the spatial peak $n_{\mathrm{peak}}(t)$ and temporal peak $t_{\mathrm{peak}}(n)$ for $\beta=\pm0.2$. The slope variations indicate changes in group velocity.
(c,d) Zoomed view of the boxed regions in (a,b), highlighting the variation in temporal intercepts $t_{\mathrm{peak}}(n)$ at the source $n=n_0$.
(e,f) Spatially normalized spatiotemporal evolution for $\beta=\pm0.2$, showing the envelope nucleation and propagation consistent with the predicted trajectories. We take $\alpha=0.1$.}}}
\raggedright
\end{figure}

\subsection*{Example: Spatial and temporal peak trajectories in a lattice with nonreciprocal stiffness and gyroscopic damping}

{{The expressions derived above allow us to interpret the spatial and temporal peak trajectories in a lattice containing both nonreciprocal stiffness and gyroscopic damping. Equation~\eqref{eq:A30} reveals that the temporal trajectory $t_{\mathrm{peak}}(n)$ consists of a standard kinematic term proportional to the inverse group velocity and a non-Hermitian temporal intercept $t_{\mathrm{int}}$. This intercept depends linearly on the temporal growth rate $\omega_I$ and inversely on the square of the real group velocity $\omega'_R$. 

Using the first-order dispersion approximations from the main text, the real group velocity is given by
\begin{equation}
\omega'_R(q) \approx \frac{k\sin q}{m\sqrt{A_0(q)}} + \frac{c\beta}{m}\cos q.
\label{eq:omegaRp_expanded}
\end{equation}
Substituting this and the expression for $\omega_I(q)$ (Eq.~\eqref{eq:combined_Im_omega_approx_expression}) into the intercept definition (Eq.~\eqref{eq:A30}) yields
\begin{equation}
t_{\mathrm{int}} \propto
\frac{
\left(
\dfrac{k\alpha\sin(q^*)}{m\sqrt{A_0(q^*)}}
- \dfrac{c_g}{2m}
\right)
}{
\left(
\dfrac{k\sin(q^*)}{m\sqrt{A_0(q^*)}}
+ \dfrac{c\beta}{m}\cos(q^*)
\right)^2 }.
\label{eq:tint_expanded}
\end{equation}
This relation makes the parameter dependencies transparent. The numerator is determined by the competition between the gain from nonreciprocal stiffness ($\alpha$) and the loss from onsite damping ($c_g$). For the parameters used in this work, the gain term dominates, ensuring that the numerator remains positive for forward propagation ($+q^*$) and negative for backward propagation ($-q^*$), consistent with the sign of the growth rate $\omega_I(q^*)$. The denominator, however, depends on the real group velocity, which is explicitly tuned by the gyroscopic damping $\beta$. Consequently, while $\alpha$ sets the sign and baseline existence of the intercept, $\beta$ modulates its magnitude through the group velocity.

These analytical trends are confirmed in Fig.~\ref{fig:FIGA1}. Panels (a,b) show that changing the sign of $\beta$ alters the slopes of both $n_{\mathrm{peak}}(t)$ and $t_{\mathrm{peak}}(n)$, consistent with the modification of $\omega'_R$ in Eq.~\eqref{eq:omegaRp_expanded}. Panels (c,d) provide enlarged views of the intercept region, highlighting the magnitude modulation predicted by Eq.~\eqref{eq:tint_expanded}. 
For fixed $\alpha=0.1$ and $q^*=0.6\pi$ (where $\cos q^* < 0$), a positive $\beta$ reduces the group velocity, while a negative $\beta$ increases it. Specifically, the velocities follow the ordering $\omega'^{(\beta=-0.2)}_{R} > \omega'^{(\beta=0)}_{R} > \omega'^{(\beta=+0.2)}_{R}$. Since the temporal intercept is inversely proportional to $(\omega'_R)^2$, the slower wave ($\beta=+0.2$) exhibits a larger intercept magnitude, while the faster wave ($\beta=-0.2$) exhibits a smaller intercept. This matches the visual evidence in panels (c,d), where $t_{\mathrm{int}}(\beta=+0.2) > t_{\mathrm{int}}(\beta=-0.2)$.
Finally, the spatiotemporal fields in panels (e,f) exhibit the corresponding asymmetric onset of the temporal envelope, confirming the analytical predictions.

In summary, the roles of the two nonreciprocal mechanisms are distinct but coupled in the temporal domain:
(i) $\alpha$ determines the sign of the temporal intercept (nucleation timing) through the growth rate $\omega_I$;
(ii) $\beta$ primarily controls the propagation speed (slope) but also provides a secondary modulation of the intercept magnitude through the group velocity.}}

\end{document}